\documentclass[fleqn,usenatbib]{mnras}
\pdfoutput=1

\usepackage{newtxtext,newtxmath}

\usepackage[T1]{fontenc}
\usepackage{ae,aecompl}

\usepackage{graphicx}	
\usepackage{amsmath}	

\usepackage{ulem}
\usepackage{url}
\usepackage{xspace}
\usepackage{comment}
\usepackage[flushleft]{threeparttable}
\graphicspath{ {./figures/} }
\newcommand{\angstrom}{\mbox{\normalfont\AA}}

\newcommand{\microns}{\ensuremath{\mu m}\xspace} 
\newcommand{\ms}{\mbox{m\,s$^{-1}$}\xspace}
\newcommand{\kms}{\mbox{km\,s$^{-1}$}\xspace}

\newcommand{\msun}{$M_{\odot}$\xspace}

\newcommand{\rsun}{$R_{\odot}$\xspace}

\newcommand{\mjup}{$M_{\rm JUP}$\xspace}

\newcommand{\mearth}{$M_\oplus$\xspace}
\newcommand{\rearth}{$R_\oplus$\xspace}

\newcommand{\Vsini}{$V \sin i$\xspace}

\newcommand{\feh}{\ensuremath{[\mbox{Fe}/\mbox{H}]}\xspace}

\newcommand{\pmra}{$\mu_{\alpha}$\xspace}
\newcommand{\pmdec}{$\mu_{\delta}$\xspace}

\newcommand{\Rmnum}[1]{\expandafter\@slowromancap\romannumeral #1@}

\newcommand{\lsun}{$L_{\odot}$}

\newcommand{\numpy}{\texttt{numpy}\xspace}
\newcommand{\matplotlib}{\texttt{matplotlib}\xspace}
\newcommand{\astropy}{\texttt{astropy}\xspace}
\newcommand{\scipy}{\texttt{scipy}\xspace}
\newcommand{\seaborn}{\texttt{seaborn}\xspace}
\newcommand{\pandas}{\texttt{pandas}\xspace}
\newcommand{\lightkurve}{\texttt{lightkurve}\xspace}
\newcommand{\isochrones}{\texttt{isochrones}\xspace}

\newcommand{\mist}{\texttt{MIST}\xspace}
\newcommand{\pytransit}{\texttt{PyTransit}\xspace}
\newcommand{\vespa}{\texttt{VESPA}\xspace}
\newcommand{\triceratops}{\texttt{TRICERATOPS}\xspace}
\newcommand{\comove}{\texttt{comove}\xspace}

\newcommand{\celerite}{\texttt{celerite}\xspace}
\newcommand{\wotan}{\texttt{wotan}\xspace}
\newcommand{\tls}{\texttt{transit-least-squares}\xspace}
\newcommand{\allesfitter}{\texttt{allesfitter}\xspace}
\newcommand{\ellc}{\texttt{ellc}\xspace}
\newcommand{\dynesty}{\texttt{dynesty}\xspace}

\newcommand{\limbdark}{\texttt{limbdark}\xspace}
\newcommand{\mrexo}{{\tt MRExo}\xspace}
\newcommand{\tqdm}{{\tt tqdm}\xspace}

\newcommand{\mstar}{\ensuremath{M_{\star}}\xspace}
\newcommand{\rstar}{\ensuremath{R_{\star}}\xspace} 
\newcommand{\lstar}{\ensuremath{L_{\star}}\xspace} 
\newcommand{\fbol}{$F_{\rm bol}$\xspace}
\newcommand{\lbol}{$L_{\rm bol}$\xspace}

\newcommand{\teff}{\ensuremath{T_{\mathrm{eff}}}\xspace}  
\newcommand{\logg}{\ensuremath{\log g}\xspace} 
\newcommand{\vsini}{\ensuremath{v \sin i}\xspace}

\newcommand{\tesspixscale}{21~arcsec/pix\xspace}
\newcommand{\spitzerpixscale}{1.2~arcsec/pix\xspace}

\newcommand{\Mp}{\ensuremath{M_{P}}\xspace} 
\newcommand{\Prot}{$P_{\rm{rot}}$\xspace}
\newcommand{\Rp}{$R_\mathrm{P}$\xspace}
\newcommand{\Porb}{$P_\mathrm{orb}$\xspace}
\newcommand{\RpRs}{$R_p/R_s$\xspace}
\newcommand{\aRs}{$a/R_s$\xspace}
\newcommand{\To}{$T_0$\xspace}

\newcommand{\Teq}{$T_{\mathrm{eq}}$\xspace}

\newcommand{\logRHK}{$\log R'_{HK}$\xspace}

\newcommand{\kepler}{\textit{Kepler}\xspace} 
\newcommand{\ktwo}{\textit{K2}\xspace}
\newcommand{\spitzer}{\textit{Spitzer}\xspace}
\newcommand{\tess}{\textit{TESS}\xspace}
\newcommand{\gaia}{\textit{Gaia}\xspace}

\newcommand{\poss}{\textit{POSS-1}\xspace}
\newcommand{\irsf}{\textit{IRSF}\xspace}
\newcommand{\ngts}{\textit{NGTS}\xspace}
\newcommand{\lco}{\textit{LCO}\xspace}
\newcommand{\pest}{\textit{PEST}\xspace}

\newcommand{\kelt}{\textit{KELT}\xspace}
\newcommand{\chiron}{\textit{CHIRON}\xspace}
\newcommand{\feros}{\textit{FEROS}\xspace}
\newcommand{\nres}{\textit{NRES}\xspace}
\newcommand{\minerva}{\textsc{Minerva}-Australis\xspace}

\newcommand{\maxrad}{\texttt{maxrad}\xspace}
\newcommand{\secthresh}{\texttt{secthresh}\xspace}

\newcommand{\target}{TOI~179\xspace}
\newcommand{\HDname}{HD~18599\xspace}
\newcommand{\HIPname}{HIP~13754\xspace}

\newcommand{\ticid}{TIC~207141131\xspace}
\newcommand{\gaiaid}{4728513943538448512\xspace}

\newcommand{\TOIdepth}{1097~ppm\xspace} 
\newcommand{\TOIperiod}{4.1374~d\xspace} 
\newcommand{\TLSperiod}{4.13~d\xspace} 
\newcommand{\TLSduration}{1.75~hr\xspace} 
\newcommand{\TLSsde}{17\xspace} 
\newcommand{\Porbval}{4.13~d\xspace}

\newcommand{\Rpval}{2.7~\rearth}
\newcommand{\Teqval}{934$\pm$10~K\xspace}
\newcommand{\MpMR}{6.66$_{-2.82}^{+15.84}$~\mearth\xspace} 
\newcommand{\MpRVfitsigma}{30.5~\mearth\xspace} 
\newcommand{\MpRVfitbest}{7.42~\mearth\xspace} 
\newcommand{\KrvMR}{2.99~\ms\xspace} 

\newcommand{\medianfilterkernelsize}{301\xspace}
\newcommand{\vespaFPP}{1.48e-11\xspace}
\newcommand{\triceratopsFPP}{0.0027$\pm$0.0008\xspace}
\newcommand{\triceratopsNFPP}{0\xspace}
\newcommand{\TSMval}{21\xspace}

\newcommand{\LiI}{Li I $\lambda$6708\xspace}
\newcommand{\LiIEWval}{42$\pm$2m\AA\xspace}
\newcommand{\ageval}{300~Myr\xspace} 
\usepackage{scalerel}
\usepackage{tikz}
\usetikzlibrary{svg.path}

\definecolor{orcidlogocol}{HTML}{A6CE39}
\tikzset{
  orcidlogo/.pic={
    \fill[orcidlogocol] svg{M256,128c0,70.7-57.3,128-128,128C57.3,256,0,198.7,0,128C0,57.3,57.3,0,128,0C198.7,0,256,57.3,256,128z};
    \fill[white] svg{M86.3,186.2H70.9V79.1h15.4v48.4V186.2z}
                 svg{M108.9,79.1h41.6c39.6,0,57,28.3,57,53.6c0,27.5-21.5,53.6-56.8,53.6h-41.8V79.1z M124.3,172.4h24.5c34.9,0,42.9-26.5,42.9-39.7c0-21.5-13.7-39.7-43.7-39.7h-23.7V172.4z}
                 svg{M88.7,56.8c0,5.5-4.5,10.1-10.1,10.1c-5.6,0-10.1-4.6-10.1-10.1c0-5.6,4.5-10.1,10.1-10.1C84.2,46.7,88.7,51.3,88.7,56.8z};
  }
}

\newcommand\orcid[1]{\href{https://orcid.org/#1}{\mbox{\scalerel*{
\begin{tikzpicture}[yscale=-1,transform shape]
\pic{orcidlogo};
\end{tikzpicture}
}{|}}}}

\title[Discovery and validation of \target\,b]{A sub-Neptune transiting the young field star \HDname at 40 pc}

\author[J.~P.~de~Leon]{\parbox{\textwidth}
{J.~P.~de~Leon\orcid{0000-0002-6424-3410}$^{1}$,
\thanks{E-mail: \texttt{jpdeleon@g.ecc.u-tokyo.ac.jp}}
J.H.~Livingston\orcid{0000-0002-4881-3620}$^{1,2,3,4}$,
J.S.~Jenkins\orcid{0000-0003-2733-8725}$^{5}$,
J.I.~Vines\orcid{0000-0002-1896-2377}$^{6}$,
R.A.~Wittenmyer\orcid{0000-0001-9957-9304}$^{7}$,
J.T.~Clark\orcid{0000-0003-3964-4658}$^{7}$,
J.I.M.~Winn\orcid{0000-0002-4265-047X}$^{8}$,
B.~Addison\orcid{0000-0003-3216-0626}$^{7}$,
S.~Ballard\orcid{0000-0002-3247-5081}$^{9}$,
D.~Bayliss\orcid{0000-0001-6023-1335}$^{10}$,
C.~Beichman\orcid{0000-0002-5627-5471}$^{11}$,
B.~Benneke\orcid{0000-0001-5578-1498}$^{12}$,
D.A.~Berardo\orcid{0000-0001-6298-412X}$^{13}$,
B.P.~Bowler$^{14}$,
T.~Brown$^{15,16}$,
E.M.~Bryant\orcid{0000-0001-7904-4441}$^{17}$,
J.~Christiansen\orcid{0000-0002-8035-4778}$^{11}$,
D.~Ciardi\orcid{0000-0002-5741-3047}$^{11}$,
K.A.~Collins\orcid{0000-0001-6588-9574}$^{18,19}$,
K.~Collins$^{19}$,
I.~Crossfield\orcid{0000-0002-1835-1891}$^{20}$,
D.~Deming\orcid{0000-0001-5727-4094}$^{21,22}$,
D.~Dragomir\orcid{0000-0003-2313-467X}$^{23}$,
C.D.~Dressing\orcid{0000-0001-8189-0233}$^{24}$,
A.~Fukui\orcid{0000-0002-4909-5763}$^{25,26}$,
T.~Gan\orcid{0000-0002-4503-9705}$^{27}$,
S.~Giacalone\orcid{0000-0002-8965-3969}$^{24}$,
S.~Gill\orcid{0000-0002-4259-0155}$^{10,28}$,
V.~Gorjian\orcid{0000-0002-8990-2101}$^{38}$,
E.~Gonz\'alez Alvarez$^{29}$,
K.~Hesse\orcid{0000-0002-2135-9018}$^{30}$,
J.~Horner\orcid{0000-0002-1160-7970}$^{7}$,
S.B.~Howell\orcid{0000-0003-0574-4853}$^{31}$,
J.M.~Jenkins$^{32}$,
S.R.~Kane\orcid{0000-0002-7084-0529}$^{33}$,
A.~Kendall$^{34}$,
J.F.~Kielkopf\orcid{0000-0003-0497-2651}$^{35}$,
L.~Kreidberg\orcid{0000-0003-0514-1147}$^{18}$,
D.W.~Latham\orcid{0000-0001-9911-7388}$^{18}$,
H.~Liu$^{36}$,
M.B.~Lund\orcid{0000-0003-2527-1598}$^{11}$,
R.~Matson\orcid{0000-0001-7233-7508}$^{31}$,
E.~Matthews\orcid{0000-0003-0593-1560}$^{13,37}$,
M.W.~Mengel\orcid{0000-0002-7830-6822}$^{7}$,
F.~Morales\orcid{0000-0001-9414-3851}$^{38}$,
M.~Mori\orcid{0000-0003-1368-6593}$^{1}$,
N.~Narita\orcid{0000-0001-8511-2981}$^{2,25,39}$,
T.~Nishiumi\orcid{0000-0003-1510-8981}$^{2,4,40}$,
J.~Okumura$^{7}$,
P.~Plavchan\orcid{0000-0002-8864-1667}$^{41}$,
S.~Quinn\orcid{0000-0002-8964-8377}$^{18}$,
M.~Rabus\orcid{0000-0003-2935-7196}$^{42,43,44}$,
G.~Ricker\orcid{0000-0003-2058-6662}$^{13}$,
A.~Rudat$^{13}$,
J.~Schlieder\orcid{0000-0001-5347-7062}$^{45}$,
R.P.~Schwarz\orcid{0000-0001-8227-1020}$^{18}$,
S.~Seager\orcid{0000-0002-6892-6948}$^{13,46,47}$,
A.~Shporer\orcid{0000-0002-1836-3120}$^{13}$,
A.M.S.~Smith$^{48}$,
A.~Sphorer\orcid{0000-0002-1836-3120}$^{13}$,
K.~Stassun\orcid{0000-0002-3481-9052}$^{49,50}$,
M.~Tamura\orcid{0000-0002-6510-0681}$^{1,2,3}$,
T.G.~Tan\orcid{0000-0001-5603-6895}$^{51}$,
C.~Tinney\orcid{0000-0002-7595-0970}$^{52}$,
R.~Vanderspek\orcid{0000-0001-6763-6562}$^{13}$,
M.W.~Werner$^{38}$,
R.G.~West$^{10,28}$,
D.~Wright\orcid{0000-0001-7294-5386}$^{7}$,
H.~Zhang$^{53}$,
G.~Zhou\orcid{0000-0002-4891-3517}$^{7,18}$
\\
{\normalsize Affiliations are listed at the end of the paper}
}
}

\date{Accepted XXX. Received YYY; in original form ZZZ}

\pubyear{2022}

\begin{document}
\label{firstpage}
\pagerange{\pageref{firstpage}--\pageref{lastpage}}
\maketitle

\begin{abstract}
Transiting exoplanets orbiting young nearby stars are ideal laboratories for testing theories of planet formation and evolution. However, to date only a handful of stars with age <1 Gyr have been found to host transiting exoplanets. Here we present the discovery and validation of a sub-Neptune around \HDname, a young (\ageval), nearby (d=40 pc) K star. 
We validate the transiting planet candidate as a \textit{bona fide} planet using data from the \tess, \spitzer, and \gaia missions, ground-based photometry from \irsf, \lco, \pest, and \ngts, speckle imaging from Gemini, and spectroscopy from \chiron, \nres, \feros, and \minerva. 
The planet has an orbital period of \Porbval, and a radius of \Rpval.
The RV data yields a 3-$\sigma$ mass upper limit of \MpRVfitsigma which is explained by either a massive companion or the large observed jitter typical for a young star. The brightness of the host star (V$\sim$9 mag) makes it conducive to detailed characterization via Doppler mass measurement which will provide a rare view into the interior structure of young planets. 

\end{abstract}


\section{Introduction}
    The majority of currently known exoplanets orbit mature host stars. In order to have a complete understanding of how planetary systems evolve from birth to maturity, we need to have a sample of systems at various evolutionary stages. In particular, young exoplanets ($<$1 Gyr in age) inhabit a very important part of the exoplanet evolutionary timescale, where formation mechanisms, accretion, migration and dynamical interactions can significantly change the shape of observed planetary systems.
    
    This is well illustrated by our knowledge of the planetary system for which we have by far the most information -- the Solar system. When we study the Solar system's history, it is immediately apparent that the first few hundred million years were an extremely chaotic and volatile time, with the system evolving dramatically on short timescales. Evidence abounds of the giant impacts that shaped both the terrestrial and giant planets, including the big smash which led to Mercury's reduced size and enhanced density \citep[e.g.][]{1988Benz, 2007Benz, 2018Chau}, the collision that shattered the proto-Earth and formed the Moon \citep[e.g.][]{1989Benz, 2012Canup, 2001Canup}, the potential origin of Mars' hemispheric anomaly \citep[e.g.][]{2008Andrews-Hanna}, the formation of Jupiter's diluted core \citep[e.g.][]{2019Liu}, and Uranus' peculiar axial tilt \citep[e.g.][]{1992Slattery, 1997Parisi, 2018Kegerreis}. At the same time, the system's small body populations reveal the scale of the migration of the giant planets. In the case of Uranus and Neptune, those small bodies reveal a significant outward migration\footnote{in the case of Neptune, travelling over at least 10 au} \citep[e.g.][]{2010bLykawka, 2011Lykawka}. 
    In the case of Jupiter and Saturn, the evidence points to significant inward migration, and potentially a period of significant dynamical instability - though the scale, rate, and chaoticity of that migration remains the subject of much debate (e.g. the Nice model: \citet{2005Tsiganis}, the Grand Tack: \citet{2014OBrien} and \citet{2012Walsh}; or a smoother, less chaotic migration: \citet{2010Lykawka} and \citet{2019Pirani}. For a detailed overview of our knowledge of the Solar system, tailored towards Exoplanetary Science, we direct the interested reader to \citet{HornerSSRev}.
    
    To date, only a handful of exoplanet host stars have a well constrained age. 
    The youngest known transiting systems so far include K2-33 \citep[age=5-10~Myr, ][]{2016DavidK2-33b, 2016MannK2-33b}, HIP~67522 \citep[17~Myr, ][]{2020RizzutoHIP67522b}, TAP~26 \citep[17~Myr, ][]{2017YuTAP26b}, Au Mic \citep[23~Myr][]{2020PlavchanAuMic}, V1298~Tau \citep[23~Myr, ][]{2019DavidV1298TauSystem}, DS~Tuc A \citep[45~Mr, ][]{2019NewtonDSTucAb}, TOI~837 \citep[45~Myr, ][]{2020BoumaTOI837b}, and TOI~942 \citep[50~Myr, ][]{2021ZhouTOI942b, 2020CarleoTOI942bc, 2021WirthTOI942b}.
    The rarity of known young planets is attributed to the difficulties involved in their detection--the large intrinsic stellar activity in young stars that induces large photometric variations and radial velocity (RV) jitter often a few orders of magnitude larger than the planet signal \citep[e.g. ][]{2021Heitzmann,2021Nicholson}. Recently, novel methods have been developed to overcome this problem by using a method to detrend both photometry and RV data \citep[e.g. ][]{2017DaiK2-131b, 2021CollierCameron}. 
    Despite the small sample of planets detected so far, there is tentative evidence that there is a measurable change in the occurrence rates of planets with time \citep[See Figure~\ref{fig:context}; ][]{2016MannK2-33b}, although more samples are needed to robustly confirm this trend. More recently, \citet{2020BergerGaiaKeplerStarsAge} and \citet{2021Sandoval} found that the ratio of super-Earth to sub-Neptune detections in the California-Kepler Survey (CKS) sample increases with system age between 1-10 Gyr even without accounting for completeness effect. Relative to sub-Neptunes, super-Earths appear to be more common around older stars despite the difficulty of detecting small planets around larger stars. Moreover, \citet{2020DavidByr} found that the size distribution of small planets also depends on the age of the planet population and that the precise location of the radius valley changes over gigayear timescales.
    However, we are still in the early stages to confidently corroborate or refute such trends given the low number of transiting planets orbiting young stars. 
    Therefore, by compiling a statistically significant sample of well-characterized exoplanets with precisely measured ages, we should be able to begin identifying new trends as well as other dominant processes governing the time-evolution of exoplanet systems.
    
    Here we present the discovery and validation of a sub-Neptune around \HDname also known as \target, a young (\ageval), nearby (d=38.6 pc) K star. 
    In Section~\ref{sec:obs}, we discuss the observations including the discovery data from \tess, follow-up photometry from space by \spitzer and from the ground by several telescopes, archival photometry and kinematics from \gaia, spectroscopy from \minerva, \feros, \chiron, and \nres, and speckle imaging from Gemini. In Section~\ref{sec:star}, we present our analyses to characterize the host star and in particular establish its youth. In Section~\ref{sec:planet}, we derive the properties of the planet such as its precise radius and mass limit. We also synthesize all available data to validate the planet candidate. In Section~\ref{sec:discussion}, we put the planet in context with the known young population. Finally, we summarize our results in Section~\ref{sec:conclusions} and also motivate further follow-up efforts of \HDname to measure the planet's mass and potentially characterize its interior and atmosphere.

\section{Observations}
    \label{sec:obs}
    \subsection{Photometry}
    \label{sec:phot}
    \subsubsection{\tess discovery} 
    \label{sec:tess}
    
\begin{figure}
        \begin{center}
        \includegraphics[clip,trim={0 10 0 0},width=\columnwidth]{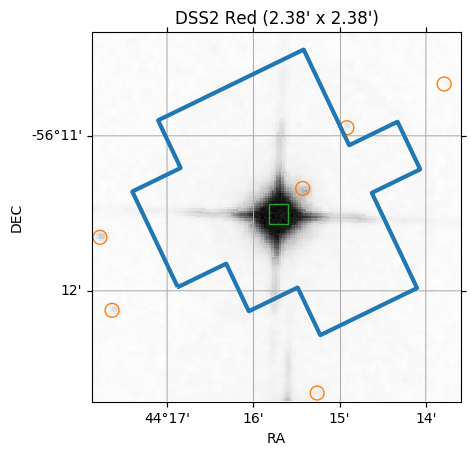}
        \caption{3" $\times$ 3" DSS2 (blue filter) image showing the target (green square) on the center and nearby \gaia sources (orange circles), superposed with the Sector 2 SPOC photometric aperture (blue polygon). We establish a dilution of <1\% for \HDname based on the flux contributions of all sources within or near the aperture perimeter. 
        \label{fig:tpf}}
        \end{center}
    \end{figure}

        \HDname was observed by \tess in sectors 2 (2018 Aug 22-Sep 2 UT; see Figure~\ref{fig:tpf}) 
        and 3 (2018 Sep 20-Oct 18 UT) during its first year with 2-minute cadence, and sectors 29 (2020 Aug 26-Sep 22 UT) and 30 (2020 Sep 22-Oct 21) during its third year with 20-sec cadence. 
        The photometric data were processed by the Science Processing Operations Center \citep[SPOC; ][]{2016JenkinsSPOC} data reduction pipeline which produced time series light curves for each campaign using Simple Aperture Photometry (SAP), and the Pre-search Data Conditioning (PDCSAP) algorithm \citep{2012StumpePDCSAPI,2012SmithPDCSAPII}. For our analysis, we used the PDCSAP light curves which have been corrected for the instrumental and systematic errors as well as light dilution effects from nearby stars. The time series in each sector has a $\sim$4-day gap 
        because of the data downlink and telescope re-pointing.
        
        The \tess object of interest (TOI) releases portal\footnote{\url{https://tess.mit.edu/toi-releases/}} announced the transiting planet candidate around \HDname (\ticid) as \target.01 \citep[]{2021GuerreroTOI}. The candidate passed all tests from the Alerts Data Validation Report \citep{2018Twicken} and is listed on the Exoplanet Follow-up Observing Program (ExoFOP)\footnote{\url{https://exofop.ipac.caltech.edu/tess/}} webpage as having a period of \TOIperiod and a transit depth of about \TOIdepth. 
        
        We performed independent transit search in the PDCSAP using \tls\footnote{\url{https://github.com/hippke/tls}} \citep{2019HippkeTLS} after applying a biweight filter using \wotan\footnote{\url{https://github.com/hippke/wotan}} with a window length of 0.5 days to detrend the stellar variability in the concatenated light curves in sectors 2 \& 3,  separately from sectors 29 \& 30 due to the long gap. A transit signal with \TLSperiod with a signal detection efficiency \citep[SDE; ][]{2016PopeC5to6} 
        of \TLSsde and a \TLSduration duration was detected. No other significant periodic transit signal was detected in further iterations.
        
    \begin{figure*}
        \begin{center}
        \includegraphics[clip,trim={0 0 0 0},width=0.85\textwidth]{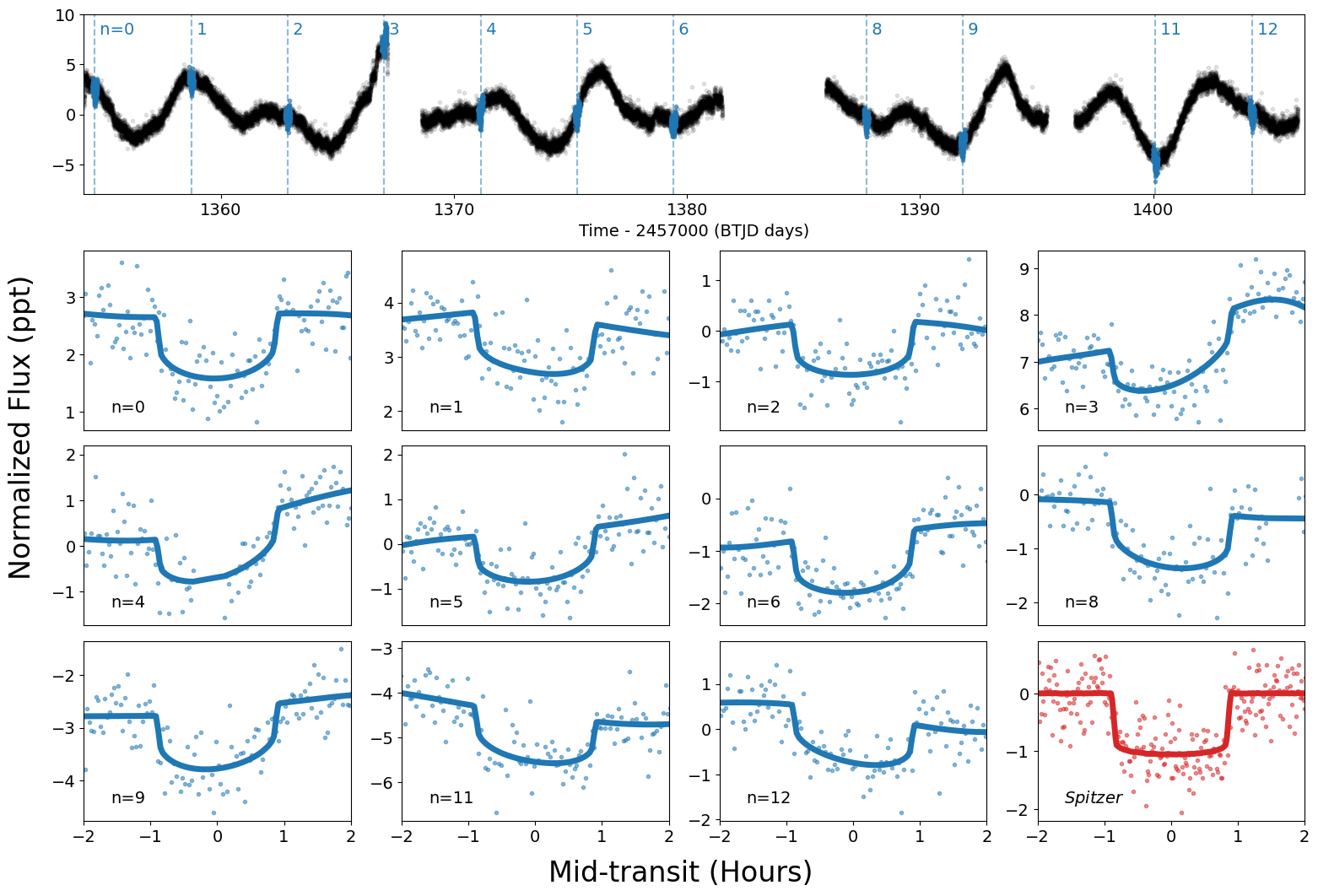}
        \caption{Raw light curves used in transit modeling superposed with best-fit transit model with baseline trend. The top row shows the \tess PDCSAP light curves (cadence=2 min) in sectors 2 \& 3 where the individual transits are highlighted in blue and numbered relative to the first (n=0). Panels 2 to 12 show the zoomed-in view of individual transits with best-fit model. The last panel shows the \spitzer light curve. \label{fig:lcs}}
        \end{center}
    \end{figure*}
    
    \begin{figure*}
        \begin{center}
        \includegraphics[clip,trim={0 0 0 0},width=0.85\textwidth]{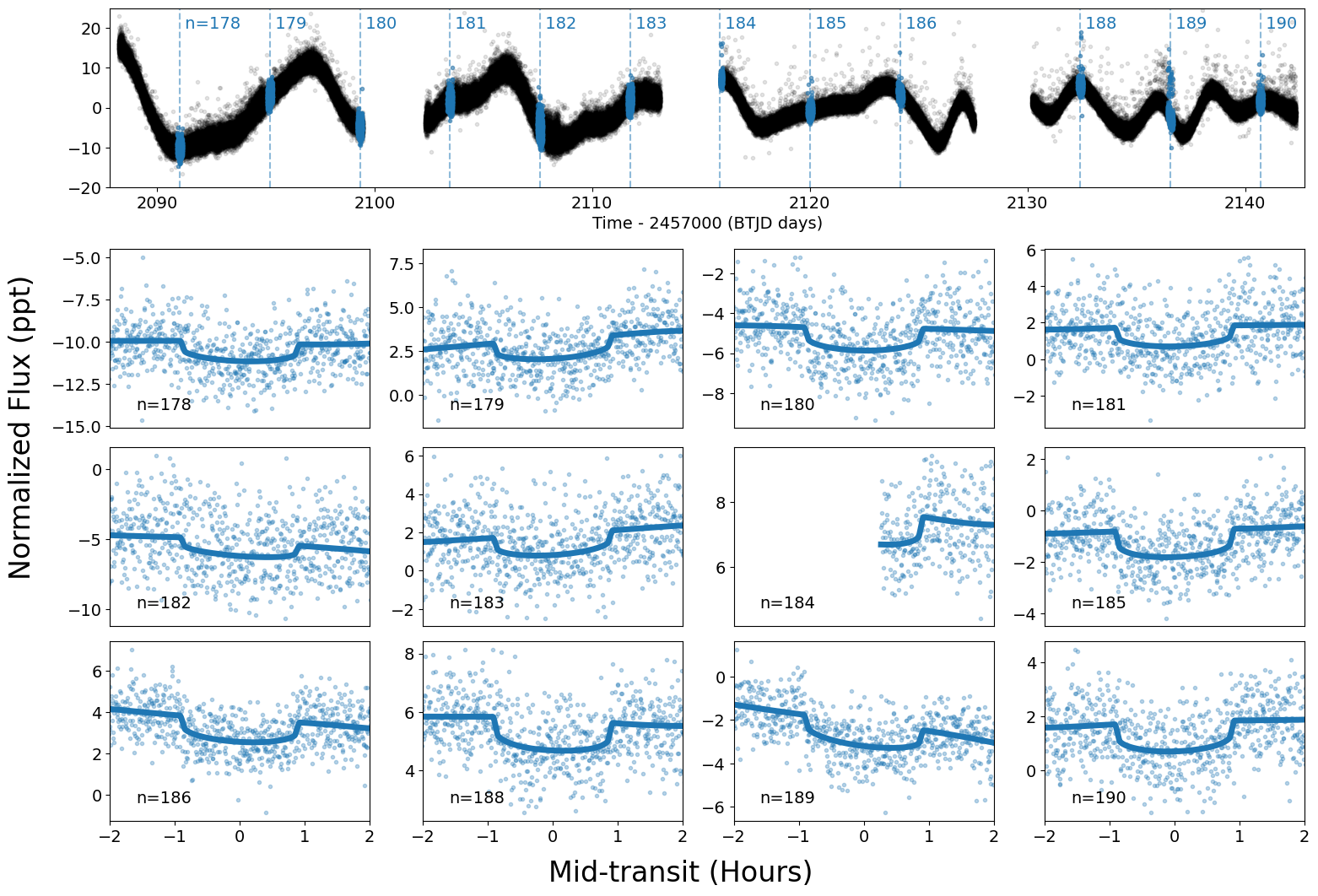}
        \caption{Same as Figure~\ref{fig:lcs} but for \tess sectors 29 and 30. Note the significantly higher sampling rate due to the shorter (20-second) cadence.
        \label{fig:lcs2}
        }
        \end{center}
    \end{figure*}
    
    \subsubsection{\spitzer follow-up} \label{sec:spitzer}
    
        We obtained observations of \HDname with the intention of following up the candidate planet using the Infrared Array Camera \citep[IRAC, ][]{1998FazioSpitzerIRAC} 4.5\microns channel, as part of \spitzer TESSTOO program 14084 (P.I. Crossfield). Observations with \spitzer have a number of advantages over those taken by \tess. First, \spitzer has a smaller pixel scale (\spitzerpixscale) than that of \tess's (\tesspixscale). This allows \spitzer observation to localize the signal by excluding the nearby stars that are blended with the target star in the \tess photometric aperture (see Figure~\ref{fig:tpf}). Second, the effects of limb darkening in \spitzer is reduced because it operates in the near-infrared, relative to \tess which operates in the optical. Third, \spitzer has better sampling of the transit due to the shorter (2-second) cadence of our \spitzer data compared to the shortest available \tess (20-second) cadence. Hence, more accurate planet parameter estimates are obtained when modeling the \spitzer transit light curve jointly with \tess. In conjunction with the \tess bandpass, the 4.5\microns IRAC bandpass also provides a relatively broad wavelength baseline which is very useful for planet validation (See Section~\ref{sec:fpp}).
        
        We used integration time of 2 seconds to keep the detector 
        from saturating and minimize data downlink bandwidth. Following \citet{2012IngallsSpitzerIRAC}, the target was placed on the "sweet spot" of the detector ideal for precise time-series photometry of bright stars like \HDname. 
        We then extracted the \spitzer light curves following \citet{2019LivingstonK2Spitzer} which was based on the approach taken by \citet{2012KnutsonHD189733} and \citet{2016Beichman}. 
        In brief, we compute aperture photometry using circular apertures centered on \HDname, for a range of radii between 2.0 and 5.0 pixels, corresponding to 2.4"--6.0". We used a step size of 0.1 pixel from 2.0 to 3.0, and a step size of 0.5 from 3.0 and 5.0. 
        The optimal aperture was selected by minimizing the photon noise due to sky background and correlated noise due to inter- and intra-pixel gain variations.
        The resulting light curve is shown in the last panel of Figure~\ref{fig:lcs}.
    
    \subsubsection{IRSF follow-up}
    
        We also conducted ground-based follow-up transit observation of \HDname on 2018 Oct. 16 using the SIRIUS camera \citep{2003NagayamaSIRIUS} on-board the 1.4-m Infrared Survey Facility (IRSF) telescope located in Sutherland, South Africa. The instrument is capable of simultaneous imaging in $J,H,K_s$ bands which is ideal for planet validation. We created light curves in three bands using the standard reduction method and aperture photometry following \citet{2013NaritaGJ1214}.
        Although, we were not able to detect the shallow 1.1 ppt (parts per thousand) event on target in all bands, we were able to rule out the deep eclipses from nearby faint stars that could reproduce the \tess detection. This adds further evidence that the signal indeed originates from \HDname as shown in Section~\ref{sec:spitzer}. 
        
    \subsubsection{\lco follow-up}
    
\begin{figure*}
            \begin{center}
            \includegraphics[clip,trim={0 0 0 0},width=\textwidth]{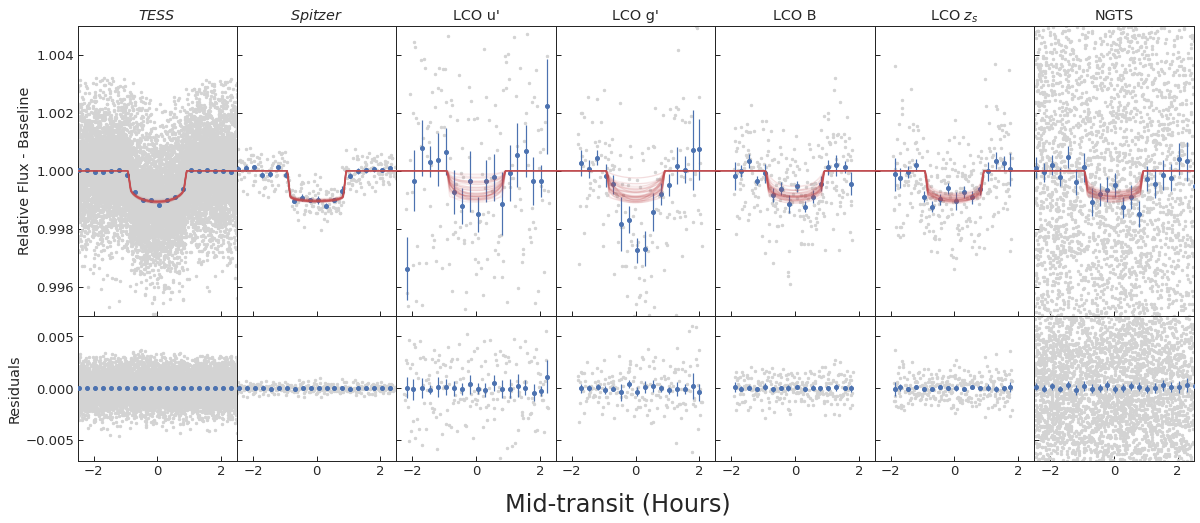}
            \caption{Light curves used in this work. The top row shows the phase-folded data with best-fit transit model (See Section~\ref{sec:transit_modeling}). The blue data points show binned data every 2-minutes and red lines show a fit using random samples from the MCMC posterior. The bottom row shows the residuals between the data and the transit+systematics model.
            }
            \label{fig:all_lcs}
            \end{center}
        \end{figure*}

        We observed \HDname on 2021 Sep 18 with 1,0m \lco-CTIO in the B and $z_s$ bands, on 2018 Nov 18 with 1.0m \lco-CPT in the $g'$ band, on 2018 Nov 23 with the 1.0m \lco-CTIO in the $u'$ band, and on 2018 Dec 22 with the 0.4m \lco-CTIO in the $i'$ band. All observations were full transit except that one taken in $i'$ band.
        We scheduled our transit observations using the \texttt{TESS Transit Finder}, which is a customized version of the Tapir software package \citep{2013Jensen}. The photometric data were calibrated and extracted using \texttt{AstroImageJ} \citep{2016CollinsAstroImageJ}. Comparison stars of similar brightness were used to produce the final light curves, each of which showed a roughly 2-ppt dip near the expected transit time. 
        The observations are summarized in Table~\ref{tab:follow-up} and plotted in Figure~\ref{fig:all_lcs}.
    
    \begin{table*}
        \centering
        \caption{Summary of our follow up photometric observations.}
        \begin{tabular}{cccccccc}
    \hline
    Telescope & Camera & Filter	& Pixel scale & Estimated PSF & Photometric Aperture & Transit  & Date (UT) \\
              &        &        & (arcsec)    & (arcsec)      & Radius (pixel)       & Coverage & \\
    \hline
     LCO-CTIO (1.0m) & Sinistro & B & 0.39 & 1.73 & 12.0 & full &  2021-09-18 \\
     LCO-CTIO (1.0m) & Sinistro & $z_s$ & 0.39 & 1.49 & 12.0 & full & 2021-09-18 \\
     MKO CDK700 (0.7m) & U16 & $i'$ & 0.401 & 5.0	& 21 & full & 2020-11-28 \\
     NGTS (0.2m) & iKon-L 936 & NGTS & 4.97 & 35 & 8.5 & full & 2019-11-21 \\
     Spitzer (0.8m) & IRAC & IRAC2: 4.5 (1.0) $\mu$m & 1.2 & 1.1 & - & full & 2019-04-29 \\
     PEST (0.3m) & ST-8XME & Ic & 1.23	& 5.3 &	7 & full & 	2019-01-07 \\
     LCO-CTIO (0.4m)	& SBIG 0.4m	&  $i'$	& 0.571	& 3.453	& 14 & partial	&	2018-12-22 \\
     LCO-CTIO (1m)	& Sinistro	&  $u'$ 	& 0.389	& 2.66	& 10 & full	&	2018-11-23 \\
     LCO-CPT (1m)	& Sinistro & $g'$ & 0.389 & - & 30 & full & 2018-11-18 \\
     IRSF (1.4m) & SIRIUS & $J,H,K$ & 0.453 & - & 10 & full & 2018-10-16 \\
     \hline
\end{tabular}
        \label{tab:follow-up}
    \end{table*}
    
    \subsubsection{\ngts follow-up}
        We observed \HDname on 2019 Nov 21 using \ngts (Next Generation Transit Survey) based at ESO's Paranal Observatory in Chile. 
        This array of twelve 0.2m telescopes is equipped with 2K$\times$2K e2V deep-depleted  Andor Ikon-L CCD cameras with 13.5$\mu$m pixels, corresponding to an on-sky size of 4.97\", each with a custom \ngts (550-927 nm) filter. These observations were performed in the 'multi telescope' observing mode \citep{2020Smith}, using three of the \ngts telescopes simultaneously. The data were reduced using a custom aperture photometry pipeline \citep{2020Bryant}, which uses the SEP library for both source extraction and photometry \citep{1996Bertin,Barbary2016}.
    
    \subsubsection{PEST followup}
        We observed \HDname in the $I_C$ band from the Perth Exoplanet Survey Telescope (PEST) near Perth, Australia. The 0.3 m telescope is equipped with a 1530$\times$1020 SBIG ST-8XME camera with an image scale of 1.2" pixel$^{-1}$ resulting in a 31'$\times$21' field of view. A custom pipeline based on \texttt{C-Munipack}\footnote{\url{https://c-munipack.sourceforge.net}} was used to calibrate the images and extract the differential photometry.
    
    \subsubsection{\kelt archival data}
    
        \HDname was observed as part of the Kilodegree Extremely Little Telescope (KELT) survey using a 42mm-aperture telescope located in Sutherland, South Africa \citep{2012PepperKELTSouth}. The telescope is equipped with Mamiya 645 80mm f/1.9 42mm lens with a 4k$\times$4k Apogee CCD, a pixel scale of $23\arcsec$ and a field of view of $26^{\circ} \times 26^{\circ}$. The target was observed with with a 20-30 minute cadence. The data were reduced in a standard manner using the pipeline described in detail in \citet{2012Siverd} and \citet{2016KuhnKELTSouth}.
    
        \begin{table}
            \centering
            \caption{\HDname photometry data.}
            \begin{tabular}{cccc}
\hline 
Time [BJD-2450000)] & Flux & $\sigma$ & Instrument\\
\hline 
8354.1116&1.0007&0.0004&\tess\\
8354.1130&1.0009&0.0004&\tess\\
8354.1157&1.0012&0.0004&\tess\\
\vdots & \vdots & \vdots & \vdots \\
\hline 
\end{tabular}
            \label{tab:phot}
            \begin{tablenotes}
                \small
                \item NOTE-- This table is published in its entirety in a machine-readable format. A few entries are shown for guidance regarding form and content.
            \end{tablenotes}
        \end{table}
        
    \subsection{Spectroscopy} \label{sec:spec}
    
        We conducted several high resolution spectroscopic observations to characterize the host star and to measure the RV variation induced by orbiting companion. In the following, we describe our observations first and then derive an upper mass limit of the companion.
    
    \subsubsection{\feros}
        We obtained 9 spectra with typical exposure time of 600 s conducted between 2019 Sep 10 and 19 UT using \feros echelle spectrograph with a resolution R=48000, wavelength coverage between 350 and 920nm, and mounted on the MPG/ESO-2.2m telescope in La Silla, Chile \citep{1999KauferFEROS}. The spectra have typical SNR of 80. The observations were performed in the Object-Calibration mode to allow precise RV observations. 
        The spectra were then processed with the CERES pipeline \citep{2017Brahm} to obtain both RVs and activity indicators. 
        The measurements are given in Table~\ref{tab:rv}, and shown in Figure~\ref{fig:rv}
    
    \subsubsection{\minerva}
    
        The MINiature Exoplanet Radial Velocity Array (\textsc{Minerva})-Australis is an observatory located in Queensland, Australia, dedicated to the precise radial-velocity and photometric follow-up of \tess planet candidates \citep[e.g. ][]{2019AdissonMINERVA, addison2021, 2022Wittenmyer} 
        It consists of four 0.7 m robotic telescopes fiber-fed to a KiwiSpec spectrograph with spectral resolution of $\sim$80,000 and wavelength coverage between 480 and 620\,nm \citep{2019WilsonMINERVA}. 
        
        We obtained 31 spectra of \HDname between 2019 Jan 6 and Jan 29 with a typical exposure time of 20-30 minutes. Radial velocities for the observations are derived for each telescope by cross-correlation, where the template being matched is the mean spectrum of each telescope. The instrumental variations are corrected by using simultaneous Thorium-Argon arc lamp observations. The measurements are given in Table~\ref{tab:rv}, and shown in Figure~\ref{fig:rv}.
    
    \subsubsection{CTIO/\chiron}
    
        We conducted high resolution spectroscopy of \HDname using CTIO High Resolution spectrometer (\chiron) on the 1.5-m SMARTS telescope. It has a spectral resolution of 80,000 with wavelength coverage between 4500 and 8900 \AA. Between 2019 Feb and 2020 Dec, we took a total of 6 spectra with typical SNR between 53 and 130. 
            
        \begin{table}
            \centering
            \caption{\HDname radial velocities.}
            \begin{tabular}{cccc}
\hline 
Time [BJD-2450000)]&RV [\ms]&$\sigma_{RV}$ [\ms]&Instrument\\
\hline 
8490.0712&-9.4571&4.3344&\minerva\\
8490.0926&-0.6974&4.1199&\minerva\\
8494.0647&21.0941&4.2761&\minerva\\
\vdots & \vdots & \vdots & \vdots \\
\hline 
\end{tabular}
            \label{tab:rv}
            \begin{tablenotes}
                \small
                \item NOTE-- This table is published in its entirety in a machine-readable format. A few entries are shown for guidance regarding form and content.
            \end{tablenotes}
        \end{table}
    
    \subsubsection{\lco/\nres}
    
        We conducted high resolution spectroscopy of \HDname using Las Cumbres Observatory's \citep[\lco; ][]{2013BrownLCO} Network of Robotic Echelle Spectrographs \citep[\nres; ][]{2018SiverdNRES}. It has a spectral resolution of 53,000 with wavelength coverage between 3800 and 8600~$\angstrom$.
        The observations were conducted on several nights with exposure times between 480s and 1200s, resulting in one observation at the unit at the Cerro-Tololo Inter-American Observatory (CTIO) in Chile and two observations at the South African Astronomical Observatory (SAAO). All three observations consisted of three consecutive exposures which were binned on a nightly basis. Due to weather conditions, only two out of the three nightly binned spectra had enough SNR for a confident spectral classification using the SpecMatch-Synth code\footnote{\url{https://github.com/petigura/specmatch-syn}}. 
        The spectrum obtained at the \lco CTIO node has SNR=102 and the spectrum obtained at the \lco SAAO node has SNR=45. 
        We note that the stellar parameters from the spectral classification are in agreement with the ones from the isochrone estimates.
    
    \subsection{\gaia astrometry} \label{sec:gaia}
    
        Between 25 July 2014 and 23 May 2016, the ESA \gaia satellite measured about 300 billion centroid positions of 1.6 billion stars. The positions, proper motions, and parallaxes of the brightest 1.3 billion sources were calculated for the second data release (DR2) \citep{2018BrownGaiaDR2}. \HDname was assigned the Gaia DR2 identifier \gaiaid, and had 265 "good" astrometric observations, indicating a nearby (d=38.6 pc), high-proper motion (-36.68, 50.60 mas/yr) star. 
        
        We further leverage \gaia DR2 to search for direct and indirect evidence of potential contaminating sources. In our sample, we can probe \gaia DR2 sources separated from the target as close as 1\arcsec.  
        \gaia DR2 can also be useful to look for hints of binarity. \citet{2018EvansGaiaBinarity} proposed that systems with large Astrometric Goodness of Fit of the astrometric solution for the source in the Along-Scan direction ($\verb|GOF_AL|>20$)) and Astrometric Excess Noise  significance ($\verb|D|>5$)\footnote{For details, see: \url{https://gea.esac.esa.int/archive/documentation/GDR2/Gaia_archive/chap_datamodel/sec_dm_main_tables/ssec_dm_gaia_source.html}} are plausibly poorly-resolved binaries. 
        Stars that are exceptionally bright or have high proper motion are proposed to explain the large offset due causing difficulties in modelling saturated or fast-moving stars, rather than unresolved binarity. We found $\verb|GOF_AL|=7.0$ and $\verb|D|=0$ for \HDname which are well below the aforementioned empirically-motivated cutoffs, indicating the target is indeed single.
    
    \subsection{Gemini speckle imaging} \label{sec:speckle}
    
\begin{figure}
            \begin{center}
            \includegraphics[clip,trim={0 0 0 0},width=\columnwidth]{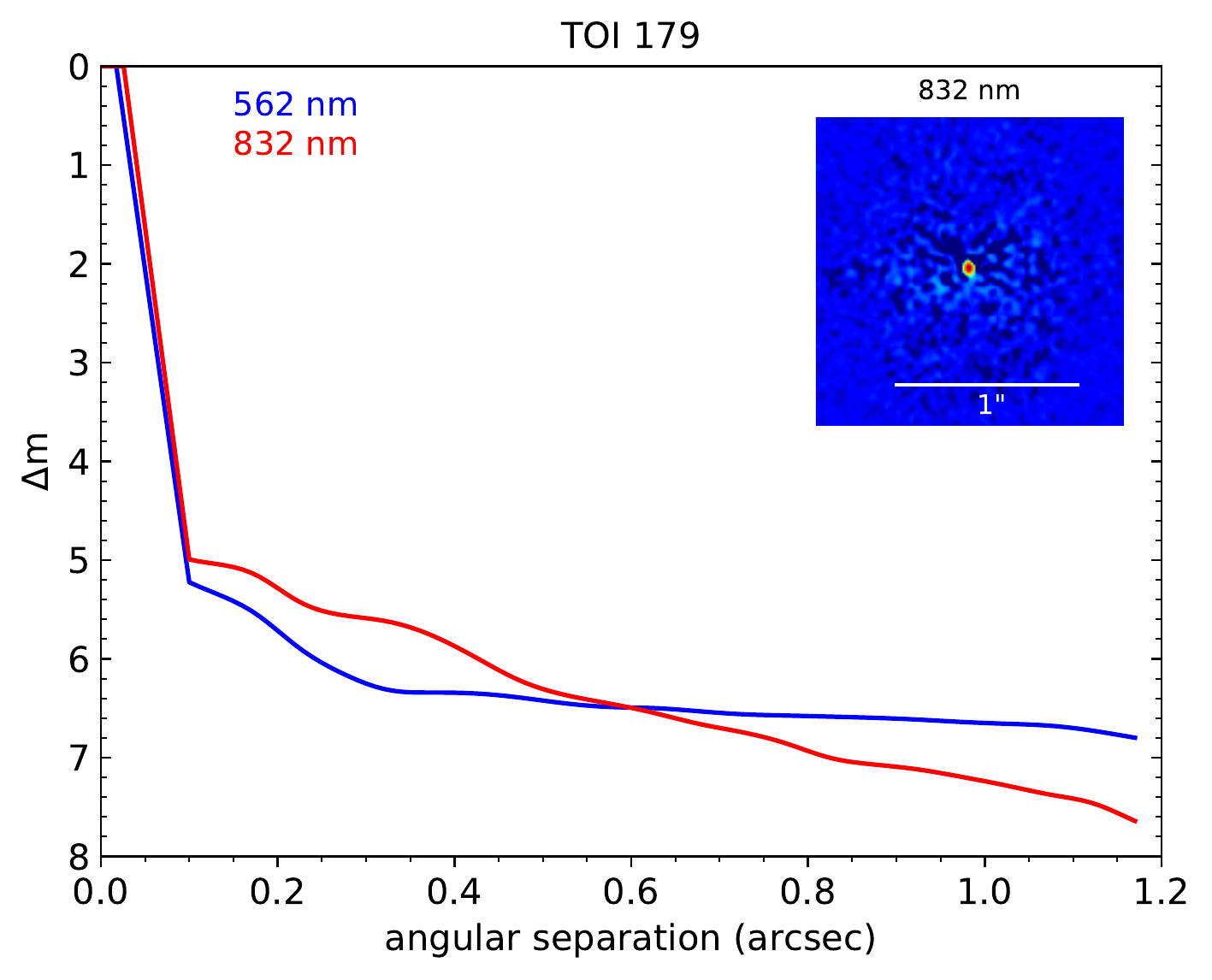}
            \caption{The contrast curves in 562 and 832nm taken by Gemini/Zorro speckle taken by the Gemini-South telescope. The inset is a reconstructed speckle image.
            \label{fig:cc}}
            \end{center}
        \end{figure}

        The presence of multiple unresolved stars in photometric and spectroscopic observations of a transiting planetary system biases measurements of the planet's radius, mass, and atmospheric conditions \citep[e.g. ][]{2020FurlanHowell, 2016SouthworthEvans}. To determine if any fainter point sources existed closer to target inside of \gaia's point-source detection limits and to rule out false positives caused by an eclipsing binary as well as to search for potential (sub-)stellar companions within a few arcseconds from the target, we conducted speckle imaging using the Zorro instrument in b and r cameras centered on 562nm and 832nm, respectively, mounted on the 8.1m Gemini South Telescope at the Cerro Pachon, Chile \citep{2021ScottZorro}. 
        Smooth contrast curves were produced from the reconstructed images by fitting a cubic spline to the 5-$\sigma$ sensitivity limits within a series of concentric annuli. The speckle observations with their corresponding contrast curves in Figure~\ref{fig:cc} illustrate that no companions were detected within a radius of 1.2\arcsec~down to a contrast level of 7 magnitudes, and no close companion star was detected within angular resolutions of the diffraction limit (0.02\arcsec) out to 1.2\arcsec. At the distance of \HDname (d=38.6 pc), these limits correspond to spatial limits of 0.8 to 46 AU.
        These observations sharply reduce the possibility that an unresolved background star is the source of the transits.
        The contrast curves are also used as additional constraints for false positive calculation in Section~\ref{sec:fpp}.

\section{Host star properties} \label{sec:star}

    \HDname (\HIPname, \target, \ticid) is a known young, nearby K2V star included in the SPHERE GTO The SpHere INfrared survey for Exoplanets (SHINE) survey sample \citep{2017Chauvin}. Based on 16 HARPS spectroscopic observations, \citet{2020GrandjeanRV,2021Grandjean} did not detect any planetary companion. They also measured a systemic RV=115.9$\pm$38.9 m/s (rms) and chromospheric flux ratio of \logRHK=-4.310 indicating a high activity index mainly due to stellar starspots. In the following, we characterize the host star in detail. 
    
    \subsection{Fundamental parameters} 
    \subsubsection{\isochrones} \label{sec:isochrones}
    To obtain the physical properties of \HDname, we utilized the Python package \isochrones \citep{2015MortonIsochrones}\footnote{http://github.com/timothydmorton/isochrones; v.2.1}
    that relies on the MESA Isochrones \& Stellar Tracks (\mist; Dotter 2016) grid to infer stellar parameters using a nested sampling scheme given photometric or spectroscopic data and other empirical constraints.
    In particular, we used 2MASS ($JHKs$) \citep{2006Skrutskie2MASS} and
    along with \gaia DR2 parallax \citep{2018BrownGaiaDR2} and extinction. 
    We corrected the parallax for the offset found in \citet{2018StassunTorresGaiaOffset} while quadratically adding 0.1 mas to the uncertainty to account for systematics in the \gaia DR2 data \citep{2018LuriGaiaUnc}.
    Additionally, we used \teff, \logg, and \feh derived from spectroscopy (see Section~\ref{sec:spec}) or taken from the literature as additional priors.
    The results of \isochrones are summarized in Table~\ref{tab:star}. 
    
    \subsubsection{Spectral Energy Distribution}

    This section presents an independent method to derive empirical stellar parameters of \HDname which also serves to cross-check our results obtained from the \isochrones method (Section~\ref{sec:isochrones}). 
    Following the procedures described in \citet{2016StassunTorres,2017Stassun,2018StassunStars}, we first construct the broadband spectral energy distribution (SED) of \HDname, and then derive the stellar radius using the \gaia DR2 parallax. The broadband photometry measurements include the FUV and NUV magnitudes from {\it GALEX}, the $B_T V_T$ magnitudes from {\it Tycho-2}, the $JHK_S$ magnitudes from {\it 2MASS}, the W1--W4 magnitudes from {\it WISE}, and the $G G_{\rm RP} G_{\rm BP}$ magnitudes from {\it Gaia}. Altogether, the available photometry spans the full stellar SED over the wavelength range 0.2--22~\microns.
    Then, we fit the SED with the NextGen stellar atmosphere models taking into account the effective temperature ($T_{\rm eff}$) and metallicity ([Fe/H]) derived from the spectroscopic analysis. The extinction ($A_V$) was set to zero because of the star being very near at d=40~pc. 
    
    \begin{figure}
        \centering
        \includegraphics[clip,trim={4cm 10 3cm 3},width=\columnwidth]{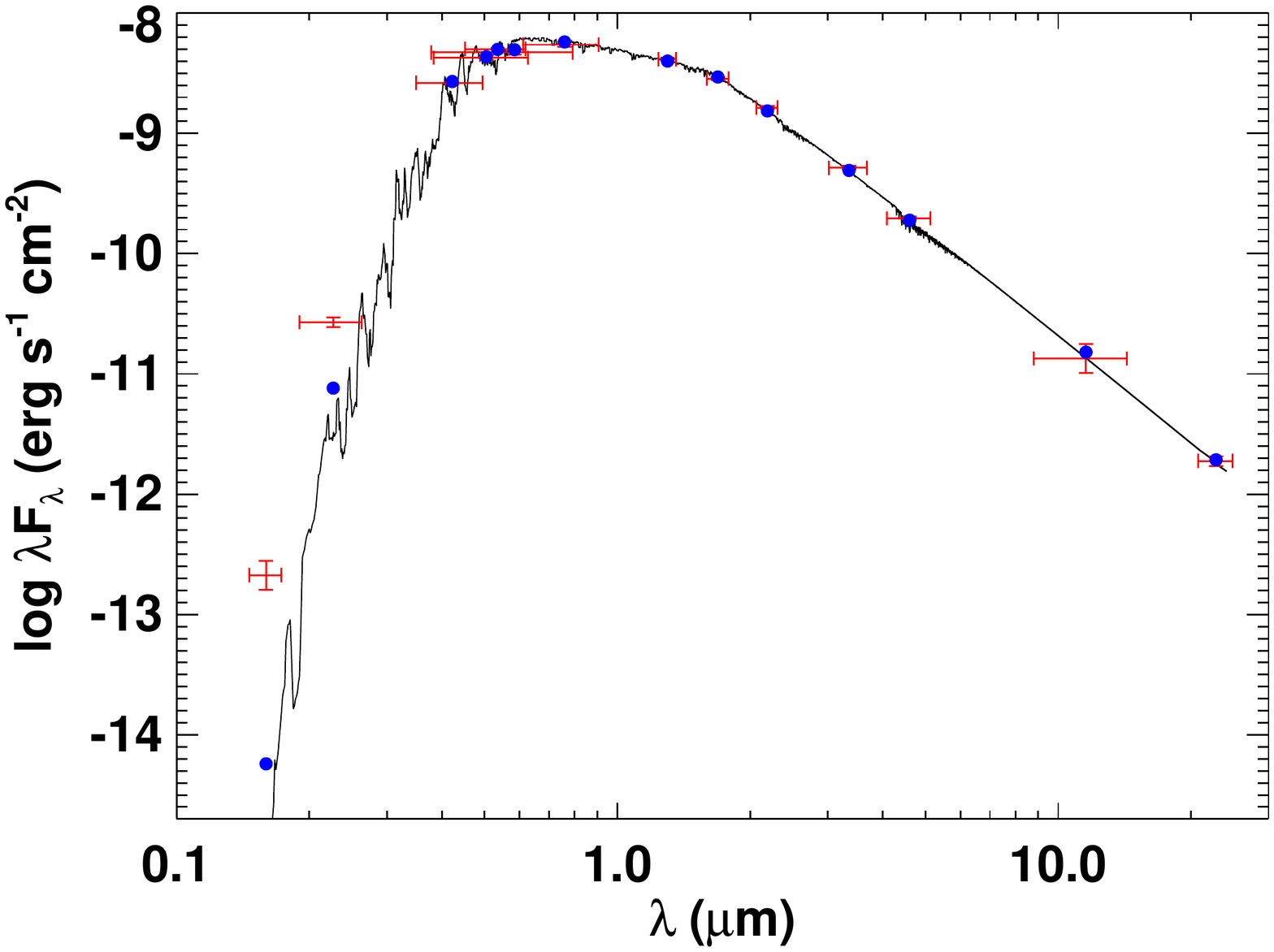}
        \caption{Spectral energy distribution (SED) of \HDname. Red symbols represent the observed photometric measurements, where the horizontal bars represent the effective width of the passband. Blue symbols are the model fluxes from the best-fit NextGen atmosphere model (black). 
        \label{fig:sed}}
    \end{figure}
    
    Figure~\ref{fig:sed} shows the SED plot with broadband photometry measurements superposed with the best-fit model. The resulting fit is excellent with a reduced $\chi^2$ of 1.9 (excluding the {\it GALEX} FUV and NUV fluxes, which are consistent with moderate chromospheric activity). Integrating the unreddened SED model yields the bolometric flux at Earth of $F_{\rm bol} = 8.013 \pm 0.093 \times 10^{-9}$ erg~s$^{-1}$~cm$^{-2}$. 
    Taking the $F_{\rm bol}$ and $T_{\rm eff}$ together with the \gaia DR2 parallax, which was adjusted to account for the systematic offset reported by \citet{2018StassunTorresGaiaOffset}, yields the stellar radius as $R_\star = 0.781 \pm 0.016$~R$_\odot$. 
    In addition, the stellar bolometric luminosity is obtained directly from $F_{\rm bol}$ and the parallax, yields $L_{\rm bol} = 0.3709 \pm 0.0044$~L$_\odot$. 
    Finally, we estimate the stellar mass from the empirical relations of \citet{2010TorresBolometricCorrection} and a 6\% error from the empirical relation itself yields $M_\star = 0.87 \pm 0.05$~$M_\odot$, whereas the mass estimated empirically from $R_\star$ together with the spectroscopic $\log g$ yields $M = 0.56 \pm 0.13$~M$_\odot$. This discrepancy suggests that the spectroscopic $\log g$ may be slightly underestimated. In any case, these values are in agreement with the values estimated using \isochrones method in Section~\ref{sec:isochrones} which uses high-resolution spectra. The final values are listed in Table~\ref{tab:star}.
    
    \begin{table}
        \small
        \centering
        \caption{Stellar parameters for \HDname.}
        \begin{tabular}{ccc}
\hline
Parameter & Value & Provenance \\
\hline
\multicolumn{3}{l}{Catalog Information}\\
R.A. (hh:mm:ss) & 02:57:02.88 & \gaia DR2 \\
Decl. (dd:mm:ss) & -56:11:30.73 & \gaia DR2 \\
\pmra (mas/yr) & -36.68$\pm$0.04 & \gaia DR2\\
\pmdec (mas/yr) & 50.60$\pm$0.05 & \gaia DR2\\
Parallax (mas) & 25.90$\pm$0.11$\dagger$  & \gaia DR2 \\
RV (km/s) & -0.484 $\pm$ 0.261 & \gaia DR2 \\ 
Distance (pc) & 38.57$\pm$0.04 & \citet{2018BailerJones} \\
TOI ID & 179 & \\
TIC ID & 207141131 & \\
HIP ID & 13754 & \\
HD ID & 18599 & \\
Gaia DR2 ID & 4728513943538448512 & \\
&&\\
\multicolumn{3}{l}{Broadband magnitudes}\\
G & 8.73$\pm$0.01 & \gaia DR2\\
B$_P$ & 9.21$\pm$0.01 & \gaia DR2\\
R$_P$ & 8.14$\pm$0.01 & \gaia DR2\\
J & 7.428$\pm$0.018 & 2MASS\\
H & 7.029$\pm$0.015 & 2MASS\\
K & 6.883$\pm$0.02 & 2MASS\\
\tess & 8.18$\pm$0.01 & TIC v8\\
&&\\
\multicolumn{3}{l}{Stellar properties from CHIRON spectra}\\
\teff (K) & 5220$\pm$50K &  This work \\
\logg (dex) & 4.6$\pm$0.1 &  This work \\
\feh (dex) & -0.1$\pm$0.1 &  This work \\
Li I EW (m\AA) & 41$\pm$7 & This work \\
\multicolumn{3}{l}{Stellar properties from NRES spectra}\\
\teff (K) & 5155 $\pm$ 100 & This work \\
\logg (dex) & 4.60 $\pm$ 0.10 & This work \\
\feh (dex)  & 0.13$\pm$0.06  & This work \\
\vsini (\kms) & 3.52 $\pm$ 0.88 & This work \\
\multicolumn{3}{l}{Stellar properties from FEROS spectra}\\
Li I EW (m\AA) & 42$\pm$2 & This work \\
\multicolumn{3}{l}{Stellar properties from SED}\\
\rstar (\rsun) & 0.781$\pm$0.016 & This work \\
\fbol (erg/s/cm$^2$) & 8.013$\pm$0.093$\times 10^{-9}$ & This work \\ 
\lbol (\lsun) & 0.3709$\pm$0.0044 & This work \\ 
\multicolumn{3}{l}{Stellar properties from \isochrones (adopted)}\\
\rstar (\rsun) & 0.78$\pm$0.01 & This work \\
\mstar (\msun) & 0.84$\pm$0.03  & This work \\
\teff (K) & 5241$\pm$44  & This work \\
\logg (dex) & 4.58$\pm$0.02  & This work \\
\feh (dex) & -0.08$\pm$0.07  & This work \\
A$_V$ (mag) & 0.21$\pm$0.14 & This work\\ 
\Vsini (\kms) & 3-9 & This work \\
\Prot (d) & 8.7 & This work \\ 
\multicolumn{3}{l}{Stellar properties from literature}\\
\mstar (\msun) & 0.91 & \citet{2021Grandjean} \\
\vsini (\kms) & 4.6 $\pm$ 0.88 & \citet{2021Grandjean} \\ 
 & 4.3 & \citet{2011Jenkins} \\
R'hk (dex) & -4.39  & \citet{2011Jenkins} \\
 & -4.4154  & \citet{2018BoroSaikia} \\
 & -4.310  & \cite{2020GrandjeanRV} \\
age (Myr) & $200^{+200}_{-75}$ & \citet{2021Grandjean} \\
\hline
\end{tabular}

        \label{tab:star}
        \begin{tablenotes}
            \small
            \item NOTE-- $^{\dagger}$0.1 mas was added in quadrature.
        \end{tablenotes}
    \end{table}
    
    \subsection{Youth indicators}
    
    \citet{2014SoderblomAges} provides a comprehensive review of the techniques to determine approximate stellar ages. 
    To age-date \HDname, first we search for coeval, phase-space neighbors and compile a sample of candidate siblings to compare with the empirical sequences of young clusters (color-magnitude diagram, $R'_{HK}$, \LiI equivalent width, stellar rotation period) and gyrochonology relationships. 
    Age dating stars is notoriously difficult except for ensemble of stars. 
    Hence, most of the currently known young stars hosting transiting planets 
    are found in stellar associations where age can be reliably measured.
    In fact, the youngest transiting host star known so far, K2-33, is found in a 5-10~Myr moving group in Upper Sco.
    According to MESA grids, the Pre-Main Sequence (PMS) contraction time of a \mstar=0.8\msun star is $\sim$67 Myr. Thus, the parameters of the star do not significantly change between 0.1 to 1 Gyr. However, we can look at various youth indicators to derive the age of \HDname. 
    
    \subsubsection{Stellar association}
        \HDname is previously known to be a young field star based on previous studies \citep{2020GrandjeanRV, 2021Grandjean}. Thus, it comes to no surprise that we did not find \HDname to be a member of any of the 1641 clusters catalogued in \citet{2020CantatGaudin} as well as among the young moving groups identified in \citep{2018GagneNYMG,2018GagneBANYANSIGMA}. 
        We double checked for co-moving stars in the neighborhood by querying the kinematics of all sources within 10' of the target from the \gaia DR2 catalog and found no match as found using the \comove\footnote{\url{https://github.com/adamkraus/Comove}} code.
        We found a match in TGv8 catalog \citep{2020CarilloTGv8} using \gaiaid and found the probability of \HDname being in the galactic thin and thick disk to be 98.5\% and 1.4\%, respectively. This is not surprising given \HDname's high galactic latitude of $\sim$53 deg.
    
    \subsubsection{Li I 6708 \AA~equivalent width}
    
    The strength of the lithium 6708\AA~absorption feature has traditionally been used as a youth indicator for Sun-like stars \citep{2014SoderblomAges}. When mixed down into a layer in a star that is roughly $2.5\times 10^6$ K, Li is burned into heavier elements, and thus rapidly destroyed within the first $\sim$500 Myr. 
    Thus, measuring the lithium 6708\AA~absorption feature is a strong evidence that the star's age is less 1 Gyr.
    We inspected \LiI in \feros and \chiron spectra to confirm the youth of \HDname. We measured an equivalent width (EW) of \LiIEWval where the uncertainty comes from the standard deviation of EW measurement for each spectra.
    This Li strength generally agrees with those found in stars in Hyades and Praesepe corresponding to an age of 800 Myr. The EW width is narrower than those measured by \citet{2019Bowler} for 58 Li-rich K to M stars (100--600 m\AA) in the solar neighborhood with ages between 10-120 Myr.
    
    \begin{figure}
        \centering
        \includegraphics[clip,trim={0 0 0 0},width=\columnwidth]{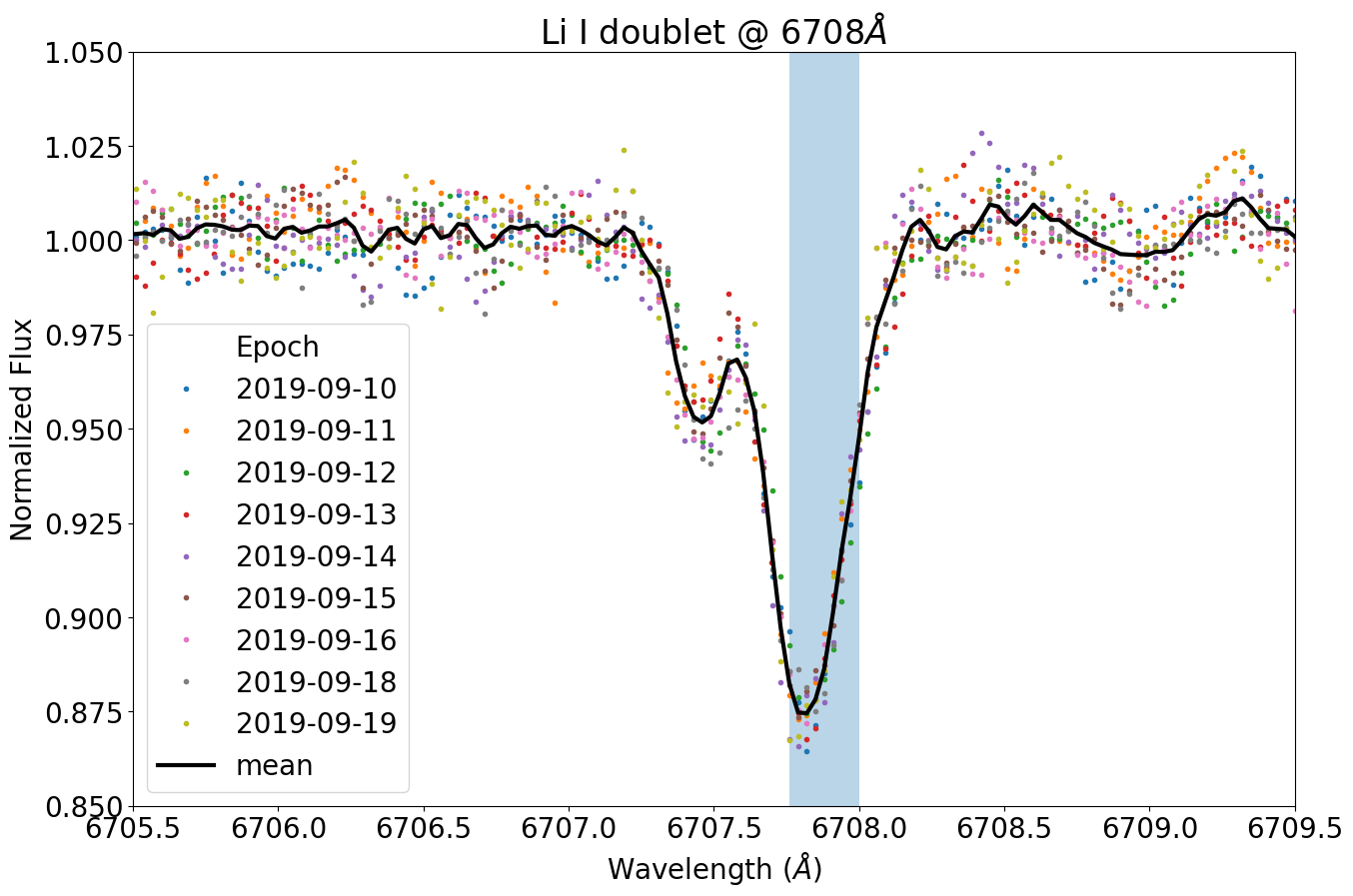}
        \caption{\feros spectra zoomed around the Li I doublet absorption at 6708 \AA~ (shaded blue region) which is a known strong indicator of youth. The black line is the mean of the all spectra taken at different epochs.}
        \label{fig:Li}
    \end{figure}
    
    \begin{figure}
        \centering
        \includegraphics[clip,trim={0 0 0 0},width=\columnwidth]{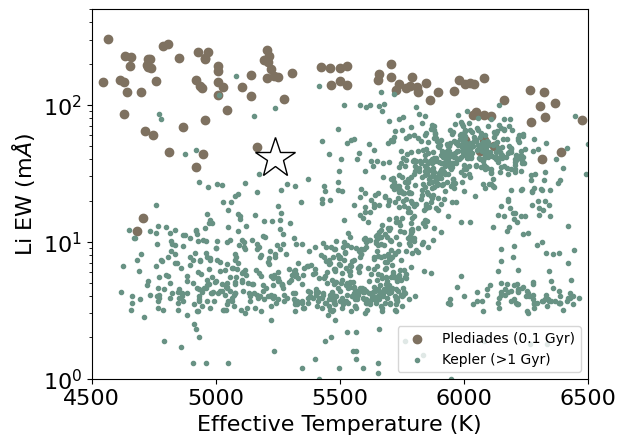}
        \caption{\teff vs. equivalent width of \HDname (masked as star) compared to Pleiades (age 0.1 Gyr) and \kepler field stars (>1 Gyr).}
        \label{fig:EW_Teff}
    \end{figure}
    
    \subsubsection{Activity indicators}
        We can also estimate the stellar age by taking advantage of the observed chromospheric activity together with empirical age-activity-rotation relations. For example, taking the chromospheric activity indicator, $\log R'_{HK} = -4.41 \pm 0.02$ from \citet{2018BoroSaikia} and applying the empirical relations of \citet{2008MamajekHillenbrand}, gives a predicted age of $0.30 \pm 0.05$~Gyr. 
        Whilst this star is an X-ray source based on detection from ROSAT, the X-ray strength is weak ($\log Lx/Lbol=-4.64\pm0.25$) which corresponds to $1-\sigma$ age range from X-ray of $475^{+734}_{-305}$ Myr. 
        The X-ray count to luminosity calibration is from \citet{1995Fleming}, and the X-ray luminosity to age calibration is from Equation A3 in \citet{2008MamajekHillenbrand}. 
    
        \subsubsection{Stellar rotation period and amplitude}
        
\begin{figure}
            \begin{tabular}{l}
                \includegraphics[clip,trim={0 180 70 0},width=\columnwidth]{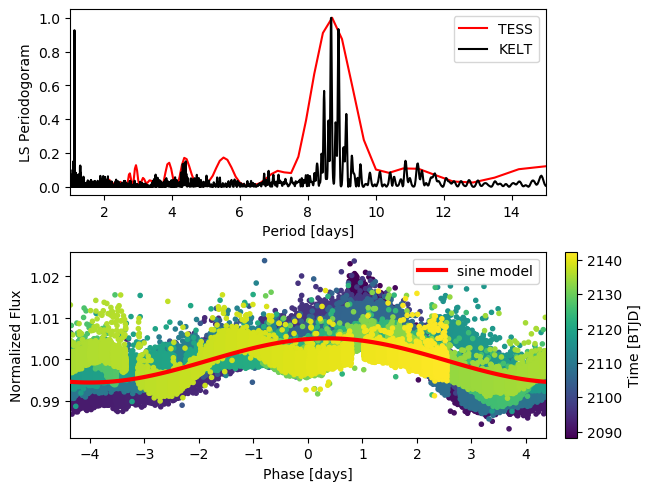}
                \\
                \includegraphics[clip,trim={0 0 0 10},width=\columnwidth]{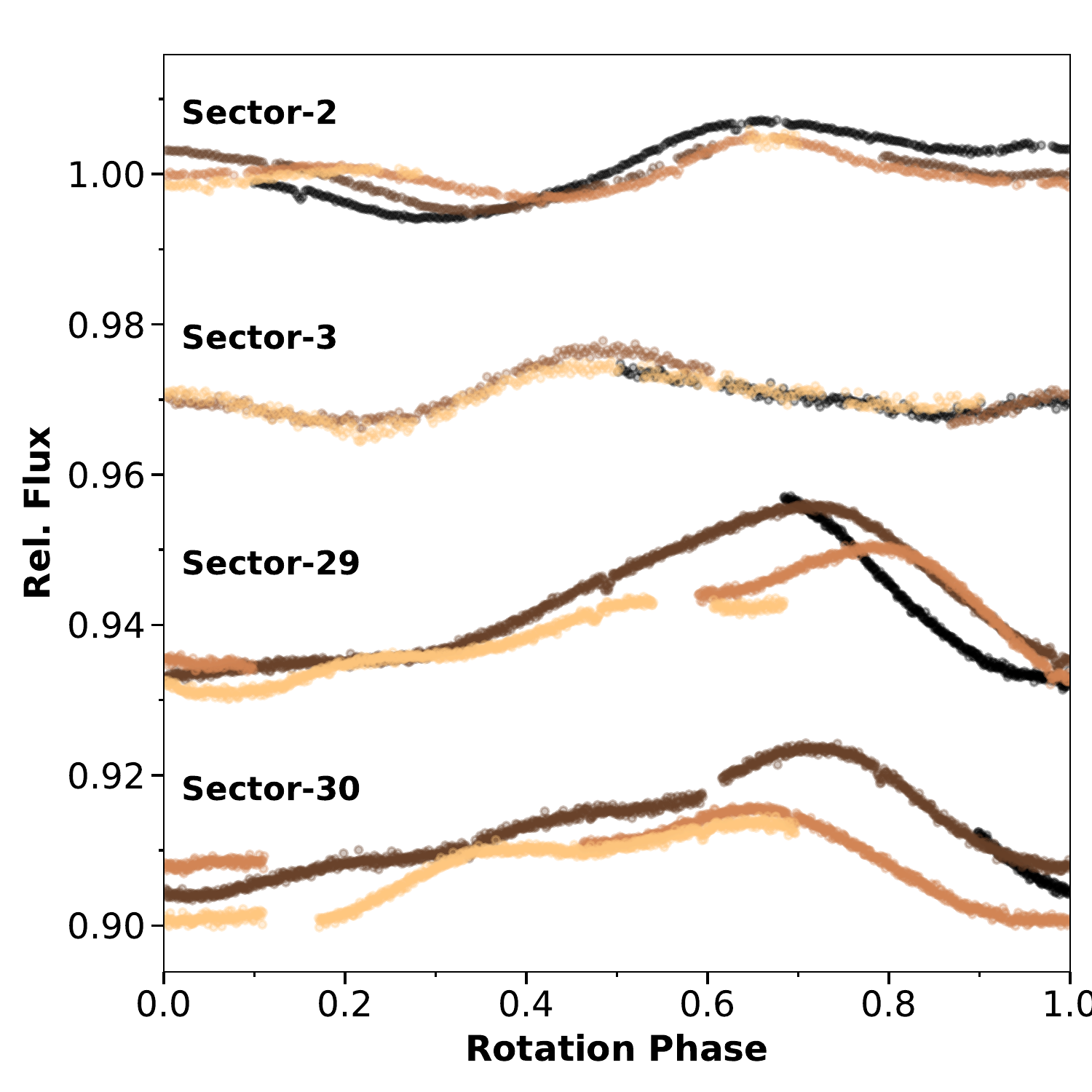}
            \end{tabular}
            \caption{Rotation period estimate. The top panel shows the Lomb-Scargle periodogram of the \tess light curves (red) and \kelt observations (black). The rotation period is consistently detected in both \tess and the six years of monitoring from \kelt. The bottom panel shows the \tess light curves folded to the rotation period. Each rotation period is over-plotted with a slightly different color gradient, where the lighter/darker color represents more recent observations.}
            \label{fig:prot_tess_kelt}
        \end{figure}

        The top panel in Figure~\ref{fig:lcs} clearly shows significant spot-modulated rotational signals in the \tess light curves. 
        To measure the rotation period of \HDname, we used the light curves from \tess PDCSAP and \kelt \citet{2018OelkersKELT}. 
        The top panel in Figure~\ref{fig:prot_tess_kelt} shows the Lomb-Scargle (LS) periodogram of 
        the four sectors of \tess (red) and the 6-year \kelt light curves (black). Both show a consistent peak at \Prot=8.8 d. 
        The bottom panel in Figure~\ref{fig:prot_tess_kelt} shows the spot evolution of \HDname over each of four \tess sectors. We note that the light curves from sectors 2 and 3 showed a strong secondary peak at 1/2 rotation period in the LS periodogram, with the corresponding phase folded light curve showing that the star likely has large spot regions on both hemispheres. By sectors 29 and 30, these spots had evolved such that the 8.71-day period is the dominant frequency in the periodogram, which we also confirmed from \kelt light curves.
        We note also that the stellar rotation period does not coincide with the planetary orbital period.
        
        We can further corroborate the activity-based age estimate by also using empirical relations to predict the stellar rotation period from the activity. For example, the empirical relation between $R'_{HK}$ and rotation period from \citet{2008MamajekHillenbrand} predicts a rotation period for this star of $9.7 \pm 1.3$~d, which is compatible with the rotation periods above as well as with the \Prot of 8.69~d reported by \kelt, and also compatible with the projected rotation period inferred from the spectroscopic $v\sin i$ and $R_\star$ which gives $P_{\rm rot}/\sin i = 7.2 \pm 3.3$~d. 
        Using \Prot=8.69 d, \rstar=0.77 \rsun, \vsini=4.3 km/s, we derive an inclination of 74 deg.
        
        \begin{figure}
            \centering
            \includegraphics[clip,trim={0 0 0 0},width=\columnwidth]{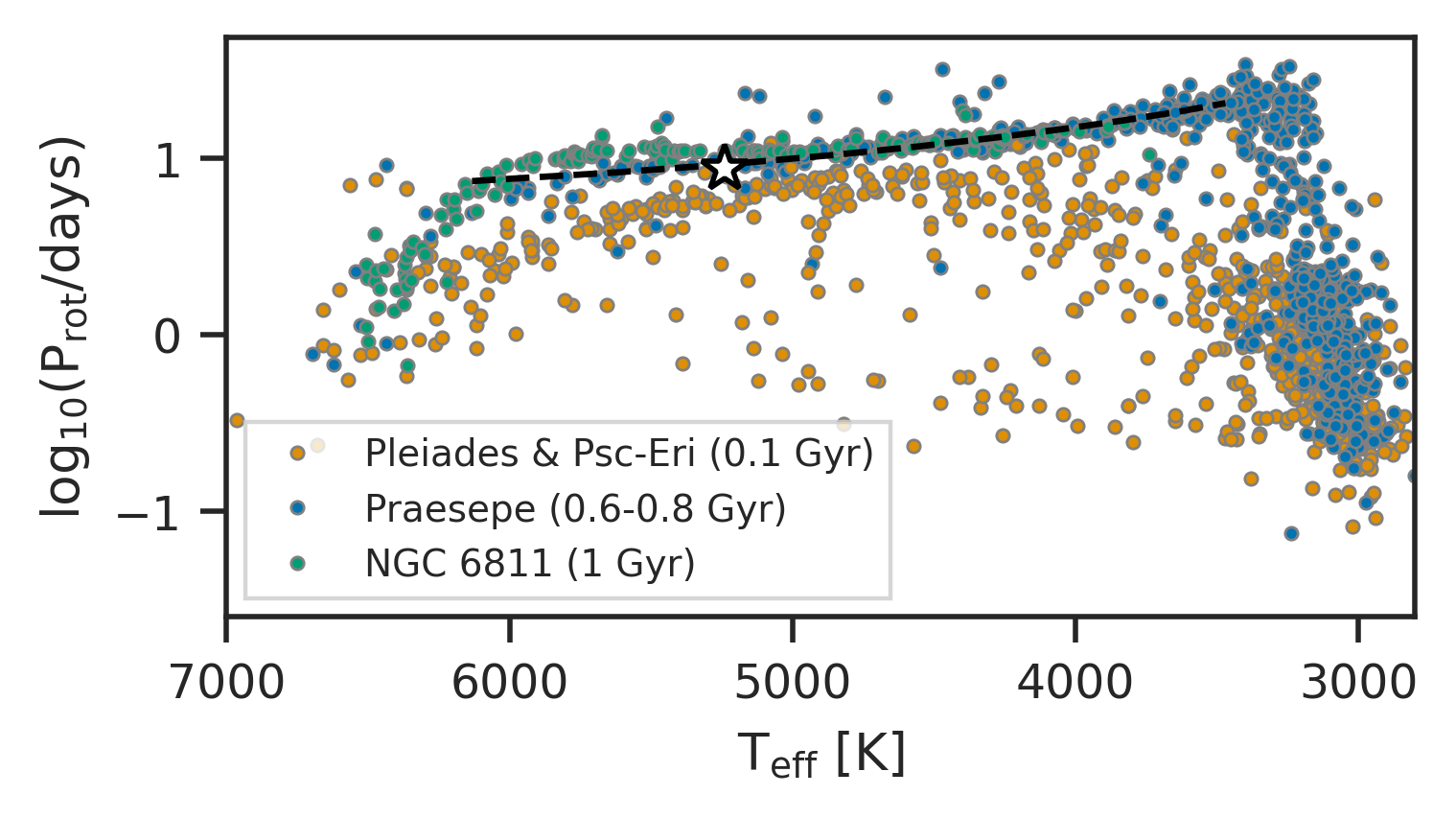}
            \caption{\teff vs. rotation period relation of \HDname (white star) compared to representative star cluster members with known ages. The black dashed line shows the negative empirical relation between rotation period and \teff.}
            \label{fig:Prot_color}
        \end{figure}
        
        Using the Generalized Lomb-Scargle (GLS) periodogram \citep{2009ZechmeisterGLS} on 
        \tess light curves, we derived \Prot and amplitude of 8.582549$\pm$0.003679~d and 0.8673$\pm$0.0014\%, respectively.
        Using the age model as a function of "smoothed" rotation amplitude presented in \citet{2020Morris}, we derived a poorly constrained age of 1145$^{+1718}_{-960}$~Myr. 
    
    \subsubsection{Gyrochronology}
    
    Finally, we can also estimate the age from the observed \Prot and empirical gyrochronology relations \citet{2008MamajekHillenbrand}, we derive a median age of 386~Myr, with 3$\sigma$ range of 261-589~Myr. Using \citet{2007Barnes}, we derive a median age of 247~Myr, with 3$\sigma$ range of 185-329~Myr. 
    
\begin{figure}
        \centering
        \includegraphics[clip,trim={0 0 0 0},width=\columnwidth]{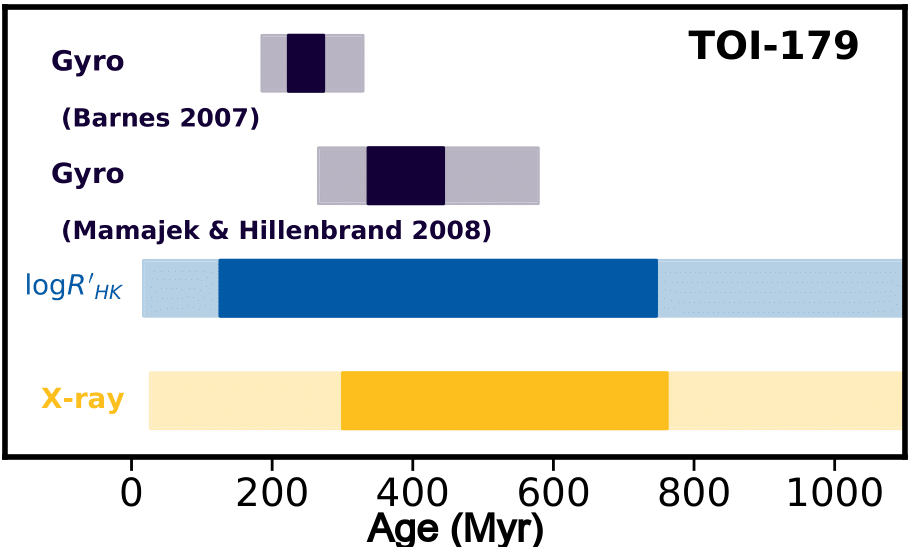}
        \caption{
        Summary of age estimates for \HDname using various methods. 
        }
        \label{fig:age_summary}
    \end{figure}

    Figure~\ref{fig:age_summary} shows the summary of the age estimates using the different methods consistent with the age between 0.1 to 1~Gyr. Assuming all the above methods are independent, we compute a weighted mean of 273$\pm$22 Myr. We adopt a median age of \ageval for the discussion in Section~\ref{sec:discussion}.
    
\section{Planet properties}\label{sec:planet}
    \subsection{Transit modeling} \label{sec:transit_modeling}
        In Section~\ref{sec:spitzer}, we confirmed that the \tess signal indeed originates from \HDname. In the following, we model the \tess and \spitzer light curves jointly to robustly 
        measure the 
        the planet's parameters, in particular the transit depth as a function of bandpass.
        
        After removing all flagged cadences and those that are more than 3-$\sigma$ above the running mean, we flattened and normalized the raw light curves using a median filter with kernel size of \medianfilterkernelsize cadences corresponding to $\sim$5$\times$ the transit duration.
        After the pre-processing steps, we model the \tess and \spitzer light curves using the Python package \allesfitter\footnote{\url{https://github.com/MNGuenther/allesfitter}} detailed in \citet{2020GuntherAllesfitter}.
        In brief, \allesfitter is a tool developed for joint modeling of photometric and RV data with flexible systematics models and was extensively used in related exoplanet studies with \tess \citep[e.g. ][]{2018HuangPiMen, 2019Dragomir,2021Addison}. Parameter estimation is done with nested sampling using \dynesty \citep{2020SpeagleDynesty} of transit and RV models defined in \ellc \citep{2016MaxtedELLC} and Gaussian Processes for systematics models using \celerite \citep{2017DFMCelerite}. 
        Our motivation to use \allesfitter is its flexible capability to model the transit and RV data either separately or jointly including model comparison. Nested sampling is also more efficient than MCMC especially for complex models that require more than 10 model parameters.
        
        We set the following as free parameters: the orbital period \Porb, mid-transit time \To, scaled semi-major axis \aRs, (cosine of) inclination $i$, 
        and quadratic limb darkening coefficients in q-space ($q_1$ and $q_2$) as prescribed by \citet{2013KippingLimbdark}. We also fit for the logarithm of the Gaussian 
        flux errors ($\log \sigma$), and the logarithm of the two hyper-parameters of the Gaussian process (GP) model with an approximated Mat\'ern-3/2 kernel for the
        out-of-transit baseline:
        \begin{equation}
        \begin{split}
            K(t)&=\sigma^2(1+1/\epsilon)\exp^{1-(1-\epsilon)\sqrt{3t}/\rho} \times \\
            & (1-1/\epsilon)\exp^{1-(1+\epsilon)\sqrt{3t}/\rho}, \epsilon=0.01,
        \end{split}
        \end{equation}
        where $\sigma$ and $\rho$ reflects the characteristic amplitude and length scale of the GP, respectively. 
        We also tried other kernels, such as simple harmonic oscillator, but Mat\'ern-3/2 kernel yielded the highest Bayesian evidence.
        
\begin{figure}
            \begin{center}
            \includegraphics[clip,trim={0 0 0 0},width=\columnwidth]{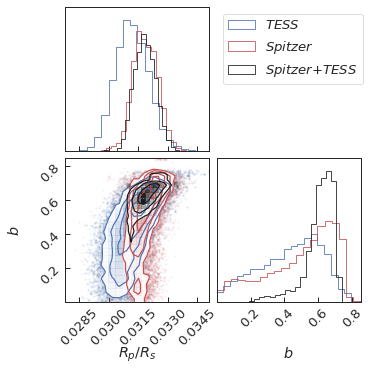}
            \caption{Comparison of marginalized posterior distributions of fits using \tess-only (blue), \spitzer-only (red), and joint (black) dataset.
            The higher radius ratio obtained from the joint fit relative to the single band fits is due to its larger and more precise impact parameter. 
            }
            \label{fig:corner}
            \end{center}
        \end{figure}

        We imposed uniform priors on the host star's density computed from the stellar parameters reported in Table~\ref{tab:star}, taking into account uncertainties. This ensures that at each step in the Nested Sampling run, the stellar density is derived from the planet's orbit (\aRs and \Porb), and the fit gets penalized for deviation from the prior.
        We also imposed (truncated) Gaussian priors on $q_1$ and $q_2$, with mean and standard deviation computed using \limbdark\footnote{\url{https://github.com/john-livingston/limbdark}}, which is a tool for Monte Carlo sampling an interpolated grid of the theoretical limb darkening coefficients (in the \tess and \spitzer bandpasses) tabulated by \citet{2012ClaretLimbDarkening} given \teff, \feh, and \logg of the host stars. We ran the nested sampler in dynamic mode with 500 live points until it reached the prescribed 1\% tolerance criterion for convergence. Based on 35611 posterior samples, we report the median and 68\% credible interval of the resulting marginalized posterior distributions in Table~\ref{tab:planet}. The \tess and \spitzer light curves with best-fit transit model is shown in Figure~\ref{fig:lcs} and Figure~\ref{fig:lcs2}. 
        The comparison of the marginalized posterior distributions of the said parameters from modeling using \tess-only, \spitzer-only, and jointly are shown in Fig.~\ref{fig:corner}. The joint transit modeling of the the \tess and \spitzer lightcurves resulted to tighter constraints on the radius ratio and impact parameter.
        
        \begin{table}
            \centering
            \caption{Results of joint transit modeling of \tess and \spitzer light curves.}
            \begin{tabular}{lc}
    \hline 
    Parameter & Value\\ 
    \hline 
    Radius ratio; $R_p / R_\star$ & $0.03209\pm0.00040$ \\ 
    Epoch; $T_{0}$ (BJD) & $2458747.64452\pm0.00024$ \\ 
    Orbital period; \Porb (d) & $4.1374386\pm0.0000029$ \\ 
    Semi-major axis over host radius; $a/R_\star$ & $13.14_{-0.23}^{+0.22}$ \\ 
    Companion radius; $R$ ($\mathrm{R_{\oplus}}$) & $2.730\pm0.049$ \\ 
    Semi-major axis; $a$ (AU) & $0.0476\pm0.0010$ \\ 
    Inclination; $i$ (deg) & $87.00\pm0.11$ \\ 
    Impact parameter; $b_\mathrm{tra}$ & $0.687\pm0.015$ \\ 
    Total transit duration; $T_\mathrm{tot}$ (h) & $1.858_{-0.012}^{+0.011}$ \\ 
    Host density from orbit; $\rho_\mathrm{\star}$ (cgs) & $2.51\pm0.13$ \\ 
    Equilibrium temperature; $T_\mathrm{eq}$ (K) & $935\pm11$ \\ 
    Transit depth (undil.); $\delta_\mathrm{tr; undil; tess}$ (ppt) & $1.078\pm0.025$ \\ 
    Transit depth (dil.); $\delta_\mathrm{tr; dil; b; tess}$ (ppt) & $1.078\pm0.025$ \\ 
    Transit depth; $\delta_\mathrm{tr; dil; b; spitzer}$ (ppt) & $1.051\pm0.027$ \\ 
    Limb darkening; $u_\mathrm{1; tess}$ & $0.54\pm0.22$ \\ 
    Limb darkening; $u_\mathrm{2; tess}$ & $0.07\pm0.22$ \\ 
    Limb darkening; $u_\mathrm{1; spitzer}$ & $0.113_{-0.069}^{+0.092}$ \\ 
    Limb darkening; $u_\mathrm{2; spitzer}$ & $0.148_{-0.092}^{+0.069}$ \\ 
    \hline
\end{tabular}
            \label{tab:planet}
        \end{table}
    
    \subsection{Companion mass constraint}
    
    To put a limit on the mass of putative \HDname\,b, we fit an RV model with a circular orbit to the RV data from \minerva and \feros based on their small RV scatter, more precise than the rest of our RV data. 
    We used the RV model included in the \pytransit python package\footnote{\url{https://github.com/hpparvi/PyTransit}} which we simplified to have five free parameters: phase-zero epoch $\rm{T_0}$, period, RV semi-amplitude, RV zero point, and RV jitter term. For the $\rm{T_0}$ and the period, we put Gaussian priors using the $\rm{T_0}$ and period derived from the transit analysis. For the other parameters we put wide uniform priors. We ran the built-in Differential Evolution optimizer and then sample the parameters with Markov Chain Monte Carlo (MCMC) using 30 walkers and 10$^4$ steps. We use the following equation from \citet{1999Cumming} to derive the planet mass,
    \begin{equation}
        \centering
        M_p = \Big( \frac{PM_s^2}{2\pi } \Big)^{1/3} \frac{K(1-e^2)^{1/2}}{\sin(i)}
    \end{equation}
    where \Mp is planet mass, \mstar is star mass, $P$ is orbital period, $K$ is RV semi-amplitude, $e$ is eccentricity (fixed to zero), and $i$ is inclination (fixed to 90$^{\circ}$).
    To propagate uncertainties, we use the posteriors for \mstar and $P$ from previous analyses. 
    
\begin{figure}
        \begin{center}
        \includegraphics[clip,trim={0 0 0 0},width=\columnwidth]{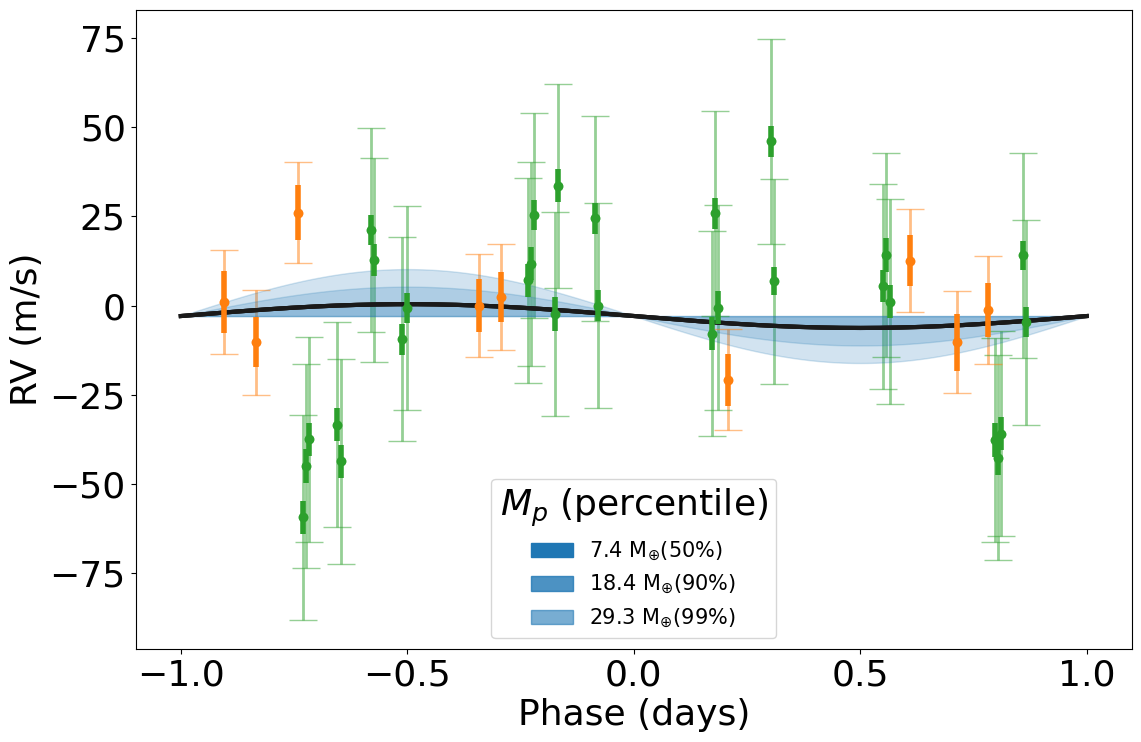}
        \caption{Phase-folded RVs with Keplerian models corresponding to the 1-, 2-, and 3-$\sigma$ mass upper limits fitted to the \minerva (green) and \feros (orange) data. Error bars with dark shades show the errors estimated from our spectroscopic analyses. The error bars with lighter shades show the original errors + jitter term value (added in quadrature) from the best-fit RV model (black line, best $\rm{M_p}=7.4$\mearth).
        \label{fig:rv}}
        \end{center}
    \end{figure}

    In Figure~\ref{fig:rv}, we plot Keplerian orbital models corresponding to different masses encompassing the 68$^\mathrm{th}$, 95$^\mathrm{th}$, and 99.7$^\mathrm{th}$ percentiles of the semi-amplitude posterior distribution. The 3-$\sigma$ upper limit is \MpRVfitsigma which places the companion 2 orders of magnitude below the deuterium burning mass limit. The best-fit semi-amplitude is \MpRVfitbest, which corresponds to a mass of $\rm{M_p}=$\MpRVfitbest, and the best-fit jitter values are $\sigma_{K,M}$=28.3~\ms and $\sigma_{K,F}$=12.4~\ms for \minerva and \feros, respectively.
    
    We calculated an expected planetary mass of \MpMR with \mrexo\footnote{\url{https://github.com/shbhuk/mrexo}}, which uses a mass-radius relationship calibrated for planets around \kepler stars \citep{2018NingMRexo}. This mass corresponds to a semi-amplitude of \KrvMR, but the observed RV data exhibits significantly larger variability ($\sigma_{RV}$=27\ms for \minerva and $\sigma_{RV}$=14\ms for \feros) as expected for young stars. We interpret this variability as being responsible for the large jitter value found by the fit, which suggests it is out-of-phase with \HDname.01. 
    \citet{2021Grandjean} also identified activity as the source of the stellar jitter.  
    We note that a detailed RV mass measurement of the companion will be presented in an accompanying paper (Vines et al., in prep.).
    
    \subsection{False positive analysis} \label{sec:fpp}
    In this section, we present the validation of the planet candidate \HDname.01. First, we describe each constraint to eliminate false positive scenarios constrained by our data. Next, we present an FPP calculation using \vespa \citep{2015MortonVespa} and \triceratops \citep{2020GiacaloneDressing} to demonstrate that the probability that \HDname.01 is an astrophysical false positive is small enough to formally validate it as a planet. 
    
    \subsubsection{Eliminating false positive scenarios} \label{sec:fpp_scenarios}
    A number of astrophysical scenarios can mimic the transit signal detected from \tess photometry, including eclipsing binary (EB) with a grazing geometry, hierarchical EB (HEB), and background/foreground EB (BEB) along the line of sight of the target. 
    Firstly, the scenario that \HDname is actually a grazing EB is ruled out based on the small (<1\kms) RV variations which disfavors stellar mass companions (Section~\ref{sec:spec}). The \tess light curves also do not exhibit secondary eclipses and odd-even transit depth variations. The light curve is also flat-bottomed as compared to V-shaped for EBs. Secondly, the scenario that \HDname is actually an NEB is ruled out due to the lack of \gaia sources found within the \tess and \spitzer photometric apertures that are bright enough to reproduce the \tess detection. 
    Thirdly, the scenario that \HDname is actually an HEB with component stars of different colors is ruled out based on the achromatic transits from \tess and \spitzer (Section~\ref{sec:transit_modeling}).
        
    The most plausible HEB scenarios for \HDname involve pairs of eclipsing M dwarfs.
    Eclipses of such stars are deeper than the K-dwarf \HDname in longer wavelengths. Limits on whether the transit depth decreases in shorter wavelengths can therefore rule out certain HEB scenarios. Similar to the procedure described in 
    \citet{2020BoumaTOI837b}, we fitted for the observed depths in different bandpasses.
    To perform the calculation, we assumed that each system was composed of the primary star (\HDname, Star 1), plus a tertiary companion (Star 3) eclipsing a secondary companion (Star 2) every \Porb d. For a grid of secondary and tertiary star masses ranging from 0.07\footnote{This mass is comparable to the lowest star mass known \citep{2017Boetticher}.} to 0.5\msun, we then calculated the observed maximum eclipse depth caused by Star 3 eclipsing Star 2 in \tess and \spitzer bandpasses using the following procedure. First, we interpolated \lstar and \teff of Star 2 and Star 3 from MIST isochrones given their masses, and the age, metallicity, and mass of Star 1 in Table~\ref{tab:star}. We then computed the blackbody functions of each stars given their \teff then convolved it with the transmission functions for each band downloaded from the SVO filter profile service\footnote{\url{http://svo2.cab.inta-csic.es/theory/fps/}}. We then integrated the result using trapezoidal method and computed the bolometric flux, \fbol, using the integrated functions above. Using Stefan-Boltzman law and given \teff and \lstar, we computed the component radii and luminosity to derive the eclipse depth. 
    
    For an HEB system with identical component stars, these stars should be very small ($m_1=m_2=$0.1\msun) taking into account dilution from the central star to reproduce the observed \tess depth ($\sim$1 ppt). 
    In this contrived scenario, the eclipse depth in \spitzer I2 is about 8 times deeper than in \tess band, because the early M-dwarf blackbody function turns over at much redder wavelengths than the central K star blackbody (Wien’s law). Thus, there is no plausible HEB configuration explored in our simulation above that can reproduce the observed depth in \spitzer I2. Even in the extreme case of a grazing orbit (i.e. $b=0.99$) for the tertiary companion, the resulting decrease in eclipse depth in \spitzer I2 is still twice as large as the observed depth. Altogether, the multiwavelength depth constraint rules out the HEB scenario. Note that the "boxy" shape of the transit signal especially in \spitzer I2 can also help rule out most of the parameter space of the HEB configurations considered above which generally produce V-shaped eclipses.
    
    Finally, the scenario that \HDname is actually a BEB is negligibly small and can be completely ruled out.
    To quantify the probability of chance alignment of a BEB to \HDname, we use the population synthesis code TRILEGAL\footnote{\url{http://stev.oapd.inaf.it/cgi-bin/trilegal}} \citep{2005GirardiTrilegal}, which simulates stellar parameters in any Galactic field. We found there is a 25\% chance to find an EB brighter than Tmag=15, within an area equal to the \tess photometric aperture (24 \tess pixels $\approx$ 2.94 arcmin$^2$). 
    We can eliminate the presence of any stellar companion with $\Delta M$<5 up exterior to 0.1" using our high spatial resolution speckle images (Section~\ref{sec:speckle})\footnote{Interior to 0.1" however, we could not clearly show that there is no bright enough star in the target's current position based on the earliest archival image taken during \poss sky survey $\sim$60 years ago when the target would have moved 0.94" relative to the background stars. The brightness of the target essentially masked the view along the line of sight to the target in the archival image}.
    This result is consistent with the observed paucity of close binaries in \tess host stars \citep{2021Ziegler, 2021Lester}. 
    Within this area, we found that there is 7.66$\times10^{-5}$ chance to find a chance-aligned star. Note that this is a conservative upper limit because this result assumes all stars are binary and preferentially oriented edge-on to produce eclipses consistent with \tess detection.
    
    Although our spectroscopic analysis disfavors the existence of an BEB spectroscopically blended with \HDname, we cannot rule out all false positive scenarios involving HEB with identical color components. For this reason, we use \vespa to model the relevant EB populations statistically. 
    
    \subsubsection{Statistical validation with \vespa and \triceratops}
    We quantify the false positive probability (FPP) of \HDname.01 using the Python package \vespa\footnote{\url{https://github.com/timothydmorton/VESPA}} which was developed as a tool for robust statistical validation of planet candidates identified by the \kepler mission \citep[e.g.][]{2012Morton} and its successor \ktwo \citep[e.g. ][]{2016CrossfieldC0to4, 2018Livingston60planets, 2018MayoC0to10}. In brief, \vespa compares the likelihood of planetary scenario to the likelihoods of several astrophysical false positive scenarios involving eclipsing binaries (EBs), hierarchical triple systems (HEBs), background eclipsing binaries (BEBs), and the double-period cases of all these scenarios. The likelihoods and priors for each scenario are based on the shape of the transit signal, the star's location in the Galaxy, and single-, binary-, and triple-star model fits to the observed photometric and spectroscopic properties of the star generated using \isochrones. 
    
    As additional constraints, we used the available speckle contrast curves described in Section~\ref{sec:speckle}, the maximum aperture radius (\maxrad)--interior to which the transit signal must be produced, and the maximum allowed depth of potential secondary eclipse (\secthresh) estimated from the given light curves. 
    Similar to \citet{2018MayoC0to10}, we computed \secthresh by binning the phase-folded light curves by the measured the transit duration and taking thrice the value of the standard deviation of the mean in each bin. Effectively, we are asserting that we did not detect a secondary eclipse at any phase (not only at phase=0.5) at 3-$\sigma$ level.
    Given these inputs, we computed a formal FPP=\vespaFPP which robustly qualifies \HDname.01 as a statistically validated planet.
    
    Additionally, we validated \HDname.01 using the Python package \triceratops\footnote{\url{https://github.com/stevengiacalone/triceratops}} which is a tool developed to validate TOIs \citep{2020GiacaloneDressing,2021GiacaloneTOI}. \triceratops validates TOIs by calculating the Bayesian probabilities of the observed transit originating from several scenarios involving the target star, nearby resolved stars, and hypothetical unresolved stars in the immediate vicinity of the target. These probabilities were then compared to calculate a false positive probability (FPP; the total probability of the transit originating from something other than a planet around target star) and a nearby false positive probability (NFPP; the total probability of the transit originating from a nearby resolved star). As an additional constraint in the analysis, we folded in the speckle imaging follow-up obtained with the Zorro imager on Gemini-South. For the sake of reliability, we performed the calculation 20 times for the planet candidate and found FPP=\triceratopsFPP and NFPP=\triceratopsNFPP (indicating that no nearby resolved stars were found to be capable of producing the observed transit). \citet{2021GiacaloneTOI} noted that TOIs with FPP < 0.015 and NFPP < 10$^{-3}$ have a high enough probability of being bona fide planets to be considered validated. Our analysis using \triceratops therefore further added evidence to the planetary nature of \HDname.01. We now refer to the planet as \HDname~b in the remaining sections.
        
    \begin{figure*}
        \centering
        \includegraphics[width=0.8\textwidth]{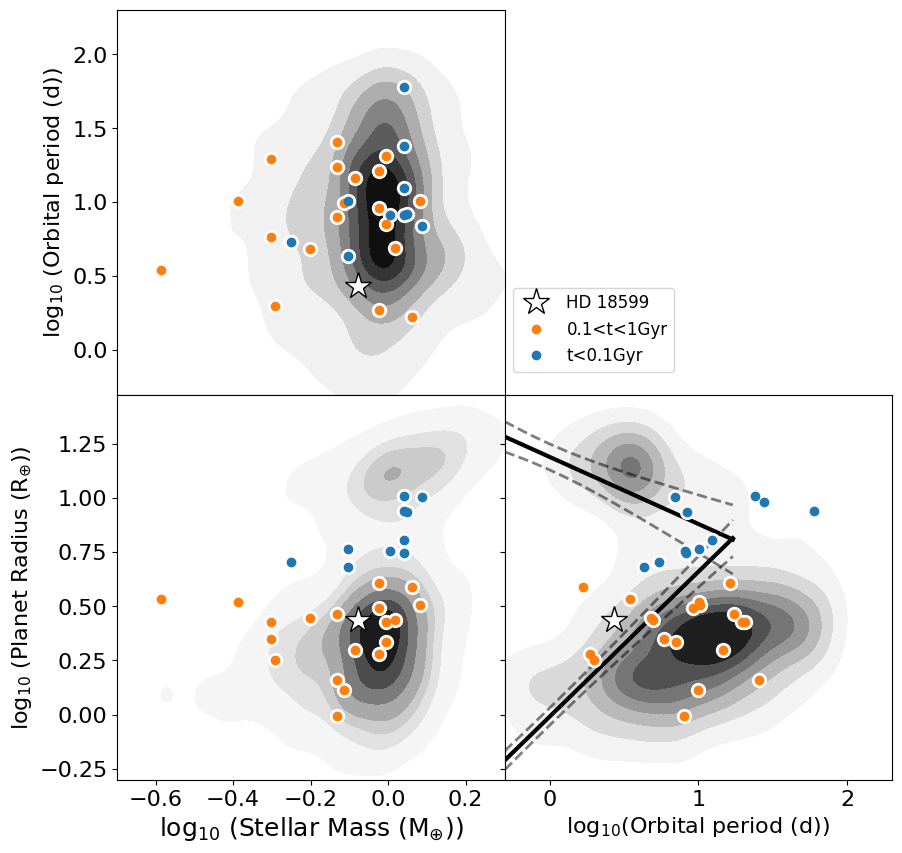}
        \caption{\HDname (marked as star) in the context of known transiting planets orbiting older field stars (gray points). Also shown in orange and blue circles are transiting planets with age<100~Myr and 100<age<1000~Myr, respectively. The contours represent the number density of the known transiting planets orbiting older host stars. \HDname\,b appears to close to the boundaries of the Neptunian desert (solid black lines, \citet{Mazeh2016}) in the period-radius plane in the lower right panel. The dashed lines refer to the boundaries' uncertainty regions.}
        \label{fig:context}
    \end{figure*}
    
\section{Discussion} \label{sec:discussion}
    
    Here, we consider the nature of \HDname\,b (=\target\,b) by placing it in context with the population of known exoplanets\footnote{Based on a query of the NASA Exoplanet Archive ``Confirmed Planets'' table on 2022 January 31, \url{https://exoplanetarchive.ipac.caltech.edu/}}.
    Figure~\ref{fig:context} shows \HDname (marked as star) in the context of known transiting planets orbiting >1~Gyr old field stars (black points). Also shown are transiting planets orbiting young (<100~Myr, blue) and adolescent stars (100<age<1000~Myr, orange). 
    \HDname~b resides in the parameter space consistent with other planets orbiting adolescent stars.
    In particular, \HDname~b is most similar to K2-284~b in terms of orbital period and radius among the known young transiting planets. K2-284~b also orbits a K-type field star with an age of 100-760~Myr, consistent with \HDname \citep{2018DavidK2-284b}.
    When compared to transiting planets around older field stars, \HDname~b appears to fall in the large-radius tail of the size distribution for close-in sub-Neptunes. 
    
    The measured period of \Porb=\Porbval and radius of \Rp=\Rpval places it close to the boundaries of the Neptunian desert defined by \citet{Mazeh2016} as shown in the lower right panel of Figure~\ref{fig:context}. \HDname~b joins the sparsely populated region within or near the boundaries of the Neptune desert, with the youngest among planet groups occupying majority of such special region. Among the known young planets, K2-100~b is located the deepest within the Neptune desert with \Porb=1.67~d and \Rp=3.88~\rearth. This is unsurprising because the planet's atmosphere is still currently experiencing photoevaporative escape \citep{2020GaidosK2100}. The youngest in the group is K2-33~b at 5-10~Myr with \Porb=5.42~d and \Rp=5.04~\rearth. 
    
    In the two lower panels in Figure~\ref{fig:context}, it is clear that the young planet population appears inflated relative to the older planet populations, especially in the case of planets orbiting low-mass stars 
    \cite[e.g. ][]{2016MannK2-33b,2020RizzutoHIP67522b,2019DavidV1298TauSystem,2019NewtonDSTucAb,2020BoumaTOI837b}.
    However, most transiting planets found orbiting adolescent stars
    like \HDname do not appear to be clear outliers in the period–radius diagram \citep[e.g. ][]{2020MannUrsaMajor,2018DavidK2-284b,2018LivingstonHyades}.
    One possible explanation for this observed behavior comes from photoevaporation theory. Young stars emit X-ray and EUV radiation strong enough to drive atmospheric escape. 
    In this picture, atmospheric loss operates in a timescale on the order of hundreds of millions of years for close-in planets with thick atmospheres to a few billion years for the largest and most massive planetary cores \citep{2020DavidByr}.
    
    Another explanation for the anomalously large radii of young planets comes from core-powered mass loss theory \citep{2019GuptaSchlichting}. In this picture, the cooling of planetary core provides the energy for atmospheric loss. This effect alone effectively reproduces the observed valley in the radius distribution of small close-in planets \citep{2017FultonGap}. 
    More recently, \citet{2020BergerGaiaKeplerStarsAge} found the first evidence of a stellar age dependence of the planet populations straddling the radius valley. They found that the fraction of super-Earths 
    to sub-Neptunes increases from 0.61 $\pm$ 0.09 at young ages (<1~Gyr) to 1.00 $\pm$ 0.10 at old ages (>1~Gyr), consistent with the prediction by core-powered mass loss which operates on a gigayear timescales. 
    
    For a mini-Neptune with a predominantly H2-He envelope presumably similar to \HDname~b, its radius will contract to approximately half of its original size in the first 500 Myr due to radiative cooling and XUV-driven mass loss \citep{2015HoweBurrows}. Given the wide age range of \HDname~b, it is difficult to ascertain whether enough time has passed to allow \HDname~b to contract to its final equilibrium radius and whether either one or combination of both effects is responsible for the observed properties of \HDname~b. At this point, we cannot be sure that the relatively large size of \HDname~b is due to its young age. However, the properties of \HDname~b do not appear to be merely a consequence of observational bias.

\section{Summary and future prospects} \label{sec:conclusions}
    Young exoplanets inhabit a very important part of the exoplanet evolutionary timescale, where formation mechanisms, accretion, migration and dynamical interactions can significantly change the shape of observed planetary systems. To date, only a handful of planetary systems <1~Gyr old are known. The main reason for the rarity of known young planets is the strong stellar activity of young stars, which makes it hard to find the subtle planetary signal in the face of large stellar variations \cite[e.g. ][]{2020PlavchanAuMic, 2019BarraganK2100}. 
    Despite the challenges, the young planet population is an emerging field that is expected to yield highly impactful scientific results. Therefore, by compiling a statistically significant sample of well-characterized exoplanets with precisely measured ages, we should be able to begin identifying the dominant processes governing the time-evolution of exoplanet systems. 
    In this light, we present the discovery and validation of a sub-Neptune orbiting the young star \HDname. Complementary to \tess data, we utilized a suite of follow-up data including photometry with \spitzer/IRAC, IRSF/SIRIUS, \& \kelt, speckle imaging with Gemini/Zorro, and high resolution spectroscopy from ESO 2.2m/\feros, \lco 2m/\nres, SMARTS 1.5m/\chiron, \& MKO 4x0.7m/. By implementing a similar validation framework and analyses presented previously, we found that the planet has an orbital period of \Porbval, a radius of \Rpval, a mass of <0.4\mjup, and an equilibrium temperature of \Teqval. 
    When compared to transiting planets around older field stars, \HDname~b appears to fall in the large-radius tail of the size distribution for close-in sub-Neptunes. 

    A comparison between the typical densities of young and old planets may be more insightful than simply comparing radii and periods. 
    However, planets around young stars are challenging for RV observations and only a handful have been successful to obtain mass measurements \citep[e.g. ][]{2019BarraganK2100,2022BarraganTOI560,2022ZicherAUMic}.
    Fortunately, the proximity and brightness of \HDname~b would allow mass measurement with high resolution Doppler spectroscopy, unlike the majority of young \ktwo planets orbiting faint host stars. 
    Measuring the planet's mass provides a rare opportunity to derive the planet's density and model its interior composition given sufficient precision. This would then allow direct comparison of bulk parameters such as density with similar sub-Neptunes orbiting older host stars.
    Our estimate of mass upper limit from Section~\ref{sec:spec} implies an RV amplitude of at most 300~\ms.
    Thus, a dedicated campaign designed to observe and model the RV systematics may allow a precise mass measurement of \HDname~b.
    A potential caveat is that \HDname is rotating faster than 15 days which results in rotational broadening at a level that limits RV precision and/or a large stellar RV jitter caused by strong stellar variability expected for young stars. 
    
    Moreover, a measurement of the planet's obliquity or the sky-projected angle between the stellar spin axis and the planet's orbital axis via observation of the Rossiter-McLaughlin (RM) effect is also a good avenue to explore. The planet's obliquity is a tracer for any dynamical processes, such as tidal interaction with stellar companions \citep[e.g. ][]{2012Batygin}, that the system might have undergone to produce a misaligned orbit, assuming well-aligned initial condition. Obliquity measurements were conducted for a growing number of young planets \citep[e.g. ][]{2022JohnsonV1298Taub, 2021WirthTOI942b, 2020GaidosK2100}. These planets were found to have well-aligned orbits which indicates that misalignments may be generated over timescales of longer than tens of Myr. 
    We estimate the RV amplitude due to the RM effect to be $\approx$140~\ms using the relation $\Delta$RV$_{\rm{RM}}$ = \vsini (\RpRs)$^2 \sqrt{1-b^2}$ \citep{2007GaudiWinnRM}, making this one of the most amenable targets for RM follow-up. This makes obliquity measurements feasible for \HDname~b, following only DS Tau Ab and AU Mic b to have obliquity measurement among young systems so far \citep{2020ZhouTOI200b,2020Martioli}. 
    Constraining the star-planet obliquity of this young system may shed light on possible migration mechanisms (e.g. Kozai, secular, tidal interactions) to explain its current architecture such as its seemingly short period as compared to other young planets \cite[e.g. ][]{2020MontetTOI200b}.
    
    Following \citet{2018Kempton}, we compute the transmission spectroscopy metric (TSM) which is a general metric useful for prioritizing of transmission spectroscopy targets for future infrared observations e.g. with JWST \citep{2020McElwainJWST}. It is defined as 
    \begin{equation}
        TSM = \alpha \times \Big( \frac{R_p T_{eq}}{M_p R_*^2} \Big) \times 10^{-M_J/5},
    \end{equation}
    where $\alpha$ is a scale factor equal to 1.28 appropriate for \HDname~b, $R_p$ is the planet radius in Earth radii, $M_p$ is the planet mass equivalent to 1.436$R_p^{1.7}$ appropriate for the size of \HDname~b, $R_*$ is the star radius in solar radii, $M_J$ is the apparent magnitude of the host star in the J band, and \Teq is the planet’s equilibrium temperature in K, 
    \begin{equation}
        T_{eq} = 0.25^{0.25} \times T_*\sqrt{\frac{R_*}{a}},
    \end{equation}
    where $a$ is the orbital semimajor axis in solar radii, and assuming zero albedo and full day-night heat redistribution. We computed TSM=\TSMval which can be interpreted as a transmission spectrum measurement with signal to noise of \TSMval in a 10 hour window with the NIRISS instrument \citep{2018LouieJWSTNIRISS}. 
    Combined with its young age, this makes \HDname a compelling target for atmospheric characterization. 
    
\section*{Acknowledgements}
We thank Blaise Kuo-tiong, Andreia Carillo, and Scarlet Saez-Elgueta for insightful discussions. This work is partly supported by JSPS KAKENHI Grant Numbers JP20K14518, JP18H05442, JP15H02063, JP22000005, and SATELLITE Research from Astrobiology Center (AB022006).
JSJ greatfully acknowledges support by FONDECYT grant 1201371 and from the ANID BASAL projects ACE210002 and FB210003.
JIV acknowledges support of CONICYT-PFCHA/Doctorado Nacional-21191829.
\minerva is supported by Australian Research Council LIEF Grant LE160100001, Discovery Grants DP180100972 and DP220100365, Mount Cuba Astronomical Foundation, and institutional partners University of Southern Queensland, UNSW Sydney, MIT, Nanjing University, George Mason University, University of Louisville, University of California Riverside, University of Florida, and The University of Texas at Austin.
D. D. acknowledges support from the TESS Guest Investigator Program grants 80NSSC21K0108 and 80NSSC22K0185.
Funding for the \tess mission is provided by NASA’s Science Mission directorate. We acknowledge the use of public \tess Alert data from pipelines at the \tess Science Office and at the \tess Science Processing Operations Center. This research has made use
of the NASA Exoplanet Archive and the Exoplanet Follow-up Observation Program website, which are operated by the California Institute of Technology, under contract with the National Aeronautics and Space Administration under the Exoplanet Exploration Program. Resources supporting this work were provided by the NASA High-End Computing (HEC) Program through the NASA Advanced Supercomputing (NAS) Division at Ames Research Center for the production of the SPOC data products. This paper includes data collected by the \tess mission, which are publicly available from the Mikulski Archive for Space Telescopes (MAST). 
We respectfully acknowledge the traditional custodians of all lands throughout Australia, and recognise their continued cultural and spiritual connection to the land, waterways, cosmos, and community. We pay our deepest respects to all Elders, ancestors and descendants of the Giabal, Jarowair, and Kambuwal nations, upon whose lands the \minerva facility at Mt Kent is situated.
This work has made use of a wide variety of public available software packages that are not referenced in the manuscript: \astropy (Astropy Collaboration et al. 2018), \scipy (Virtanen et al. 2019), \numpy (Oliphant 2006), \matplotlib (Hunter 2007), \tqdm (da Costa-Luis 2019), \pandas (The pandas development team 2020), \seaborn (Waskom et al. 2020), and \lightkurve (Lightkurve Collaboration et al. 2018).

\section*{Data availability}
    The data underlying this article were accessed from MAST (\url{https://archive.stsci.edu/hlsp/}) with specific links mentioned in the article. 
    The tables presented in this work will also be made available at the CDS (\url{http://cdsarc.u-strasbg.fr/}).
    


\bibliographystyle{mnras}
\bibliography{ref} 

Affiliations \newline
$^{1}$Department of Astronomy, Graduate School of Science, The University of Tokyo, 7-3-1 Hongo, Bunkyo-ku, Tokyo 113-0033, Japan\\
$^{2}$Astrobiology Center, 2-21-1 Osawa, Mitaka, Tokyo 181-8588, Japan\\
$^{3}$National Astronomical Observatory of Japan, 2-21-1 Osawa, Mitaka, Tokyo 181-8588, Japan\\
$^{4}$Department of Astronomical Science, The Graduated University for Advanced Studies, SOKENDAI, 2-21-1, Osawa, Mitaka, Tokyo, 181-8588, Japan\\
$^{5}$N\'ucleo de Astronom\'ia, Facultad de Ingenier\'ia y Ciencias, Universidad Diego Portales, Av. Ej\'ercito 441, Santiago, Chile Centro de Astrof\'isica y Tecnolog\'ias Afines (CATA), Casilla 36-D, Santiago, Chile\\
$^{6}$Universidad de Chile, Camino el Observatorio, 1515, Las Condes, Santiago, Chile\\
$^{7}$University of Southern Queensland, Centre for Astrophysics, West Street, Toowoomba, QLD 4350 Australia\\
$^{8}$Department of Astrophysical Sciences, Princeton University, 4 Ivy Lane, Princeton, NJ 08544, USA\\
$^{9}$Department of Astronomy, University of Florida, 211 Bryant Space Science Center, Gainesville, FL, 32611, USA\\
$^{10}$Department of Physics, University of Warwick, Gibbet Hill Road, Coventry CV4 7AL, UK\\
$^{11}$NASA Exoplanet Science Institute, California Institute of Technology, Pasadena, CA 91106, USA\\
$^{12}$Department of Physics and Institute for Research on Exoplanets, Universit{' e} de Montr{' e}al, Montr{' e}al, QC, Canada\\
$^{13}$Department of Physics and Kavli Institute for Astrophysics and Space Research, Massachusetts Institute of Technology, Cambridge, MA 02139, USA\\
$^{14}$Department of Astronomy, The University of Texas at Austin, TX 78712, USA\\
$^{15}$CASA, University of Colorado, Boulder, CO 80309, USA\\
$^{16}$Las Cumbres Observatory, Goleta CA 93117, USA\\
$^{17}$Mullard Space Science Laboratory, University College London, Holmbury St Mary, Dorking, Surrey, RH5 6NT, UK\\
$^{18}$Center for Astrophysics | Harvard \& Smithsonian, 60 Garden Street, Cambridge, MA 02138, USA\\
$^{19}$George Mason University, 4400 University Drive, Fairfax, VA, 22030, USA\\
$^{20}$Department of Physics and Astronomy, University of Kansas, 1082 Malott, 1251 Wescoe Hall Dr. Lawrence, KS 66045, USA\\
$^{21}$NASA Astrobiology Institute's Virtual Planetary Laboratory, USA\\
$^{22}$Department of Astronomy, University of Maryland, College Park, MD 20742, USA\\
$^{23}$Department of Physics and Astronomy, University of New Mexico, 210 Yale Blvd NE, Albuquerque, NM 87106, USA\\
$^{24}$Department of Astronomy, University of California, Berkeley, Berkeley, CA, USA\\
$^{25}$Komaba Institute for Science, The University of Tokyo, 3-8-1 Komaba, Meguro, Tokyo 153-8902, Japan\\
$^{26}$Instituto de Astrof{' i}sica de Canarias, V{' i}a L{' a}ctea s/n, E-38205 La Laguna, Tenerife, Spain\\
$^{27}$Department of Astronomy and Tsinghua Centre for Astrophysics, Tsinghua University, Beijing 100084, China\\
$^{28}$Centre for Exoplanets and Habitability, University of Warwick, Gibbet Hill Road, Coventry CV4 7AL, UK\\
$^{29}$Centro de Astrobiolog{' i}a (CSIC-INTA), Carretera de Ajalvir km 4, E-28850 Torrej{' o}n de Ardoz, Madrid, Spain\\
$^{30}$Wesleyan University, Middletown, CT 06459, USA\\
$^{31}$NASA Ames Research Center, Moffett Field, CA 94035, USA\\
$^{32}$Intelligent Systems Division, NASA Ames Research Center, Moffett Field, CA 94035, USA\\
$^{33}$Department of Earth and Planetary Sciences, University of California, Riverside, CA 92521, USA\\
$^{34}$School of Physics and Astronomy, University of Leicester, Leicester LE1 7RH, UK\\
$^{35}$Department of Physics and Astronomy, University of Louisville, Louisville, KY 40292, USA\\
$^{36}$School of Astronomy and Space Science, Key Laboratory of Modern Astronomy and Astrophysics in Ministry of Education, Nanjing University, Nanjing 210046, Jiangsu, China\\
$^{37}$Observatoire de Gen{' e}ve, Chemin Pegasi 51, 1290 Versoix, Switzerland\\
$^{38}$NASA Jet Propulsion Laboratory, Pasadena, CA, USA\\
$^{39}$Instituto de Astrof{' i}sica de Canarias, 38205 La Laguna, Tenerife, Spain\\
$^{40}$Department of Multi-Disciplinary Sciences, Graduate School of Arts and Sciences, The University of Tokyo, 3-8-1 Komaba, Meguro, Tokyo 153-8902, Japan\\
$^{41}$George Mason University, 4400 University Drive MS 3F3, Fairfax, VA 22030, USA\\
$^{42}$Las Cumbres Observatory Global Telescope, 6740 Cortona Dr., Suite 102, Goleta, CA 93111, USA\\
$^{43}$Department of Physics, University of California, Santa Barbara, CA 93106-9530, USA\\
$^{44}$Departamento de Matem{' a}tica y F{' i}sica Aplicadas, Universidad Cat{' o}lica de la Sant{' i}Ă‚Â­sima Concepci{' o}n, Alonso de Rivera 2850, Concepci{' o}n, Chile\\
$^{45}$NASA Goddard Space Flight Center, 8800 Greenbelt Rd, Greenbelt, MD 20771, USA\\
$^{46}$Department of Earth, Atmospheric and Planetary Sciences, Massachusetts Institute of Technology, Cambridge, MA 02139, USA\\
$^{47}$Department of Aeronautics and Astronautics, MIT, 77 Massachusetts Avenue, Cambridge, MA 02139, USA\\
$^{48}$Institute of Planetary Research, German Aerospace Center (DLR), Rutherfordstr. 2, 12489 Berlin, Germany\\
$^{49}$Department of Physics and Astronomy, Vanderbilt University, VU Station 1807, Nashville, TN 37235, USA\\
$^{50}$Department of Physics, Fisk University, Nashville, TN 37208, USA\\
$^{51}$Perth Exoplanet Survey Telescope, Perth, Western Australia\\
$^{52}$Exoplanetary Science at UNSW, School of Physics, UNSW Sydney, NSW 2052, Australia\\
$^{53}$Shanghai Astronomical Observatory, Chinese Academy of Sciences, Shanghai 200030, China\\


\bsp	
\label{lastpage}
\end{document}